\documentclass[sn-basic]{sn-jnl}

\usepackage[T1]{fontenc}%
\usepackage{xurl}%
\usepackage{graphicx}%
\usepackage{multirow}%
\usepackage{amsmath,amssymb,amsfonts}%
\usepackage{amsthm}%
\usepackage{mathrsfs}%
\usepackage[title]{appendix}%
\usepackage{xcolor}%
\usepackage{textcomp}%
\usepackage{manyfoot}%
\usepackage{booktabs}%
\usepackage{algorithm}%
\usepackage{algorithmicx}%
\usepackage{algpseudocode}%
\usepackage{listings}%
\usepackage{lscape}%
\usepackage{adjustbox}
\usepackage{anyfontsize}
\usepackage{natbib}

\begin{document}

\title[Tabular Image: a method to convert tabular data to images for convolutional neural networks]{Tabular Image: a method to convert tabular data to images for convolutional neural networks}


\author[1]{\fnm{Junhao} \sur{Liang}}

\author*[1]{\fnm{Xingjie} \sur{Wei}}\email{x.wei1@leeds.ac.uk} 

\author[1]{\fnm{Barbara} \sur{Summers}}

\affil[1]{\orgdiv{Centre for Decision Research}, \orgname{Leeds University Business School, University of Leeds}, \orgaddress{\city{Leeds}, \postcode{LS2 9JT}, \state{England}, \country{United Kingdom}}}


\abstract{Improving the predictive capability of credit scoring models is always an active area of research in the financial sector. Recognising the impressive effectiveness of neural networks in different domains (such as computer vision and natural language processing), various neural networks have been tested to potentially improve loan default prediction on credit data. Nevertheless, a significant challenge emerges due to the predominantly tabular nature of credit data, which is not well-suited to the structure and strengths of neural networks, hindering their ability to surpass traditional machine learning models in credit scoring. To overcome the challenge, we propose a novel data transformation method called \textit{Tabular Image} that converts tabular data into images to take advantage of the powerful two-dimensional convolutional neural networks that perform extremely well on images while mitigating the challenges tabular data poses to deep networks. The \textit{Tabular Image} can convert tabular data into compact and resilient images compared with existing transformation methods by creatively embedding two crucial measures in credit scoring, the weight of evidence and information value, in the image. Applications to three credit scoring benchmark datasets suggest that simply training a two-dimensional convolutional neural network with \textit{Tabular Image} can provide state-of-the-art predictive performance. In addition, the advantage of our proposed method's prediction is more evident in the large dataset. Our innovative approach raises the possibility of leveraging two-dimensional convolutional neural networks in credit scoring using a proper data representation method. Furthermore, a flexible framework is provided to suit various tabular datasets in other domains.}




\maketitle

\section{Introduction}
\label{sec:intro}

Credit risk management holds a prominent position in most financial institutions to mitigate loan losses and optimise profit. Historically, financial institutions separated default and non-default credit applicants by credit managers' intuitive experience to control credit risk \citep{lewis_an_1992}. However, as the total loans grew and the volume of loan applications increased, credit managers needed a fast and accurate way to identify default credit applicants because even a fraction of percent of increment in default rate may relate to a significant amount of loan losses \citep{baesens_benchmarking_2003,henley_construction_1997,west_neural_2000}. As a result, credit scoring was proposed to separate default and non-default credit applicants by using prediction models to convert credit applicants' financial information into a score representing credit applicants' creditworthiness \citep{lewis_an_1992}. \citet{durand_review_1942} was the first to use statistical models to separate default and non-default credit applicants. After that, various statistical and machine learning models such as logistic regression \citep{dumitrescu_machine_2022,wiginton_note_1980}, random forest \citep{brown_experimental_2012,wang_two_2012}, support vector machine \citep{harris_credit_2015,huang_credit_2007}, and gradient boosting decision trees \citep{chang_application_2018,gunnarsson_deep_2021,lessmann_benchmarking_2015} were applied to identify more accurate credit scoring models to achieve even small improvements in prediction accuracy. Despite the application of various statistical and machine learning models, it is still challenging to identify a consistently superior approach for credit scoring tasks \citep{dastile_statistical_2020,lessmann_benchmarking_2015}, so developing more accurate credit scoring models continues to be one of the crucial goals of credit scoring research.

To meet this goal, an emerging area in credit scoring research is applying neural networks to predict the probability of default. Because data used in credit scoring is usually tabular data represented in one-dimensional (1D) format, previous studies have tended to focus on MLP-like architectures \citep{baesens_benchmarking_2003,elhosenyDeepLearningBasedModel2022,hamori_ensemble_2018,lessmann_benchmarking_2015,west_neural_2000,zhaoInvestigationImprovementMultilayer2015} and 1D convolutional neural networks (CNNs) \citep{kvamme_predicting_2018,huangImprovingFinancialDistress2023, qianSoftReorderingOnedimensional2023} that accept 1D format data as input. Although recent studies have explored the possibility of using neural networks in credit scoring, the application of deep networks is still a challenge because of the characteristics of tabular data, such as mixed feature types (numerical, ordinal, and categorical), data sparsity (missing values), and lack of robustness to uninformative features \citep{grinsztajnWhyTreebasedModels2022a,shwartz-ziv_tabular_2022}. In addition, MLPs and 1D CNNs suffer from the vanishing gradient problem \citep{gilesLearningExtractingFinite1992}, thus making adding hidden layers to boost performance difficult. These challenges led to the result that eXtreme Gradient Boosting (XGBoost) was usually demonstrated to outperform neural networks for tabular data in credit scoring and raise the necessity of exploring the possibility of applying other novel neural networks \citep{gunnarsson_deep_2021}. 

So far, due to the nature of the tabular data, little attention has been paid to the more advanced, well-developed, and powerful deep learning networks such as two-dimensional (2D) CNNs. Compared to MLPs and 1D CNNs, 2D CNNs achieved impressive results or even outperformed human experts in computer vision, recognition and prediction, gaming, art imitation, etc. \citep{abdel-hamid_convolutional_2014,karpathy_largescale_2014,krizhevsky_imagenet_2017,lecun_deep_2015,mahbobiCreditRiskClassification2023,yuan_spectrumbased_2017}. Furthermore, 2D Convolutional Neural Networks (CNNs) have an inherent capability to capture high-level features automatically through stacking deep convolutional, pooling, and activation layers. This aligns with the process of feature engineering, which is a crucial and time-consuming step in building credit scoring. 2D CNNs can automate this process, minimising the need for manual feature engineering that otherwise relies heavily on expert knowledge and is resource-intensive.

This paper proposed a novel method to convert tabular data into images to utilise powerful 2D CNNs that perform exceptionally well on homogeneous data (such as images and videos) while mitigating the challenges neural networks meet when applied to tabular data. We call our method \textit{Tabular Image}, which transforms tabular data into images based on the weight of evidence (WOE) \citep{siddiqi_credit_2012} and information value (IV) \citep{hand_good_2005}. The results demonstrated that the proposed \textit{Tabular Image} training with a 2D CNN model performed better than its shallower counterpart, 1D CNN and outperformed most prediction models trained on tabular data, indicating the potential of \textit{Tabular Image} and the power of 2D CNNs. Our proposed method also outperformed 2D CNNs with other tabular data-image transformation methods, showing \textit{Tabular Image} can aid 2D CNNs in extracting signals in data, thus further boosting the prediction performance. Meanwhile, transforming tabular data into images with Tabular Image can enhance human-computer interaction. It provides an intuitive understanding of complicated tabular data, thus helping credit managers gain insight into data and identify suspicious loan applications.

The remainder of this paper is organised as follows. Section 2 reviews the deep learning models used in credit scoring and techniques used to transform tabular data into images. Section 3 describes the detailed processes of converting tabular data into images. Section 4 describes the data preparation process, the details of models and the evaluation metrics used in this study. Section 5 presents the results. The discussion is presented in section 6. We then provide a conclusion with some future perspectives in section 7.
 
\section{Related works}
\label{sec:lit}

\subsection{Deep neural networks in credit scoring}
\label{subsec:dnn}

As computing power and the volume of data continue to grow, the interest in applying neural networks to credit scoring tasks increases. Extensive research has been conducted regarding the application of fully connected neural networks in credit scoring. Among various neural networks, multilayer perceptron (MLP), restricted Boltzmann machine (RBM), and deep belief neural networks (DBN) have been mainly used to construct credit scoring models \citep{dastile_statistical_2020,gunnarsson_deep_2021}. For example, \citet{blanco_credit_2013} fitted 14 MLP credit scoring models and compared these models with linear discriminant analysis, quadratic discriminant analysis, and logistic regression. The result showed that MLP credit scoring models provided higher accuracy and lower misclassification costs than traditional models. \citet{tomczak_classification_2015} applied classification RBMs to construct an explainable scoring table and provided high prediction performance. \citet{luo_deep_2017} applied deep belief networks, which consist of a stack of RBMs, to construct a corporate credit scoring model. The prediction performance of the DBN provided the best performance compared with logistic regression, MLP and support vector machine. 

Besides fully connected neural networks, deep learning architectures that demonstrated success in natural language processing (NLP) and computer vision (CV) have been investigated, with the expectation of shedding some light on improving the prediction accuracy of credit scoring models. \citet{wangDeepLearningApproach2019a} and \citet{shen_new_2021} applied Long Short-Term Memory (LSTM), a deep learning model commonly used in NLP, on transaction data to improve the prediction accuracy. By combining LSTM with data balancing techniques and attention mechanisms, their results showed a noticeable improvement compared to traditional credit scoring models. Apart from the success of recurrent neural networks, CNNs have also been tested in the credit scoring domain. \citet{kvamme_predicting_2018} used CNNs to predict mortgage default and provided high prediction performance on transaction data. \citet{he_novel_2021} constructed a CNN as a feature generation method and constructed an ensemble model for default prediction, which significantly improved prediction performance. Although these studies showed promising results in applying neural networks to credit scoring, the prediction performance of neural networks in credit scoring is debatable. 

On the one hand, some studies present evidence that neural networks can surpass the performance of traditional models in credit scoring. \citet{west_neural_2000} compared five neural networks with five quantitative models and claimed that neural networks achieve better performance than five statistical and machine learning models in credit scoring tasks. \citet{yu_social_2015} conducted a comprehensive review of the social credit literature and pointed out that neural networks outperform statistical models in credit risk detection tasks. \citet{dastile_statistical_2020} systematically reviewed 74 articles ranging from 2010 to 2018 and claimed that neural networks perform better than statistical and machine learning models.

On the other hand, some literature suggests that the advantages of neural networks are not always clear-cut. \citet{baesens_benchmarking_2003} compared the performance of various statistical, traditional machine learning and deep learning models and concluded that the performance of logistic regression was not statistically different from neural networks. \citet{lessmann_benchmarking_2015} compared 41 classifiers on eight credit scoring datasets and observed that the prediction performance of random forest outperformed neural networks. \citet{gunnarsson_deep_2021} compared MLP and DBN with logistic regression, decision tree, random forest and XGBoost. The results showed that neural networks did not outperform machine learning models, and XGBoost was the best method among the models tested in this study.

\subsection{Data transformation}
\label{subsec:dataTrans}

Given the ongoing debate in the literature, it is evident that further investigation is needed to better understand and enhance the applications of neural networks in credit scoring. One way to utilise neural networks is to transform the tabular data into a more homogeneous format \citep{borisovDeepNeuralNetworks2022}. By implementing this type of transformation, researchers expect to be able to apply neural networks, such as 2D CNNs, which perform extremely well for classification tasks on homogeneous data. To the best of our knowledge, there is very limited research on data transformation in credit scoring. \citet{hosakaBankruptcyPredictionUsing2019} proposed a data transformation method to utilise financial ratios extracted from a company's financial statements by transforming these ratios into a grayscale image to predict bankruptcy. While this transformation method can convert tabular data into images and utilise 2D CNNs, it does not accept categorical features as input, which is required in other application areas such as individual credit default prediction. Furthermore, financial ratios are features mainly used for company bankruptcy prediction, making it difficult to extend this method to individual credit default prediction datasets. \citet{zhu_hybrid_2018} proposed a hybrid model using a relief algorithm as a feature selection tool and a 2D CNN to predict the probability of default. This study transformed tabular data into grey images to utilise 2D CNNs. This study first applied discretisation to numerical features to convert them into categorical features. It then reshaped features into binary vectors using one-hot encoding \citep{hancockSurveyCategoricalData2020} and combined all features into a sparse binary matrix, which can be considered a pixel matrix. The result showed that the Relief-CNN yielded better prediction performance than random forest and logistic regression. Similarly, \citet{dastile_making_2021} presented attempts at applying a 2D CNN to credit scoring by transforming numerical and categorical features into grey images. This study first discretised numerical features and then converted both discretised numerical features and categorical features into a pixel matrix using one-hot encoding. Although these two studies applied data transformation on tabular data, the transformation technique they used, one-hot encoding, may exacerbate the "curse of dimensionality" problem and create high-dimensional sparse feature vectors which are composed of a large number of pixels not containing information \citep{borisovDeepNeuralNetworks2022}. As a result, a significant proportion of the image may be blank. Also, the transformation method may generate large images when there is a considerable number of features or a large number of categories in categorical features, which may require a large amount of computer resources to train the CNN.

In contrast, data transformation is widely used in other fields. \citet{sharma_deepinsight_2019} proposed the DeepInsight method to transform RNA-seq data into images by projecting the high dimensional data to a 2D space using feature similarity measuring techniques and dimensionality reduction techniques, and its results outperformed those of the random forest. \citet{bazgir_representation_2020} proposed the REFINE method to convert unorganised tabular data into images based on the similarity between features calculated by a Bayesian metric multidimensional scaling approach. The results demonstrated that the method provided better predictive accuracy than conventional models. \citet{zhu_converting_2021} proposed the IGTD method for converting tabular data into compact images by assigning features to pixels based on the difference in pairwise distance rankings between features and assigned pixels. The result showed that the IGTD method performed better than DeepInsight and REFINE on drug screening datasets. Although methods proposed by \citet{sharma_deepinsight_2019}, \citet{bazgir_representation_2020}, and \citet{zhu_converting_2021} produced promising results, these methods were designed with the assumption that data with strong feature similarities, such as RNA sequence, gene or drug data, would be used, which may not be suitable for credit scoring datasets.  \citet{sun_supertml_2019} proposed the superTML method to project features in the tabular data onto black-and-white images and applied the method to four popular datasets available on the UCI Machine Learning Repository and the Kaggle platform. By projecting features in the tabular data onto images, CNNs are able to learn the shape of numbers and extract nonlinear features in the images.

\section{Proposed Methodology}
\label{sec:method}
\begin{figure}[h!]
    \centering
    \includegraphics[scale=0.15]{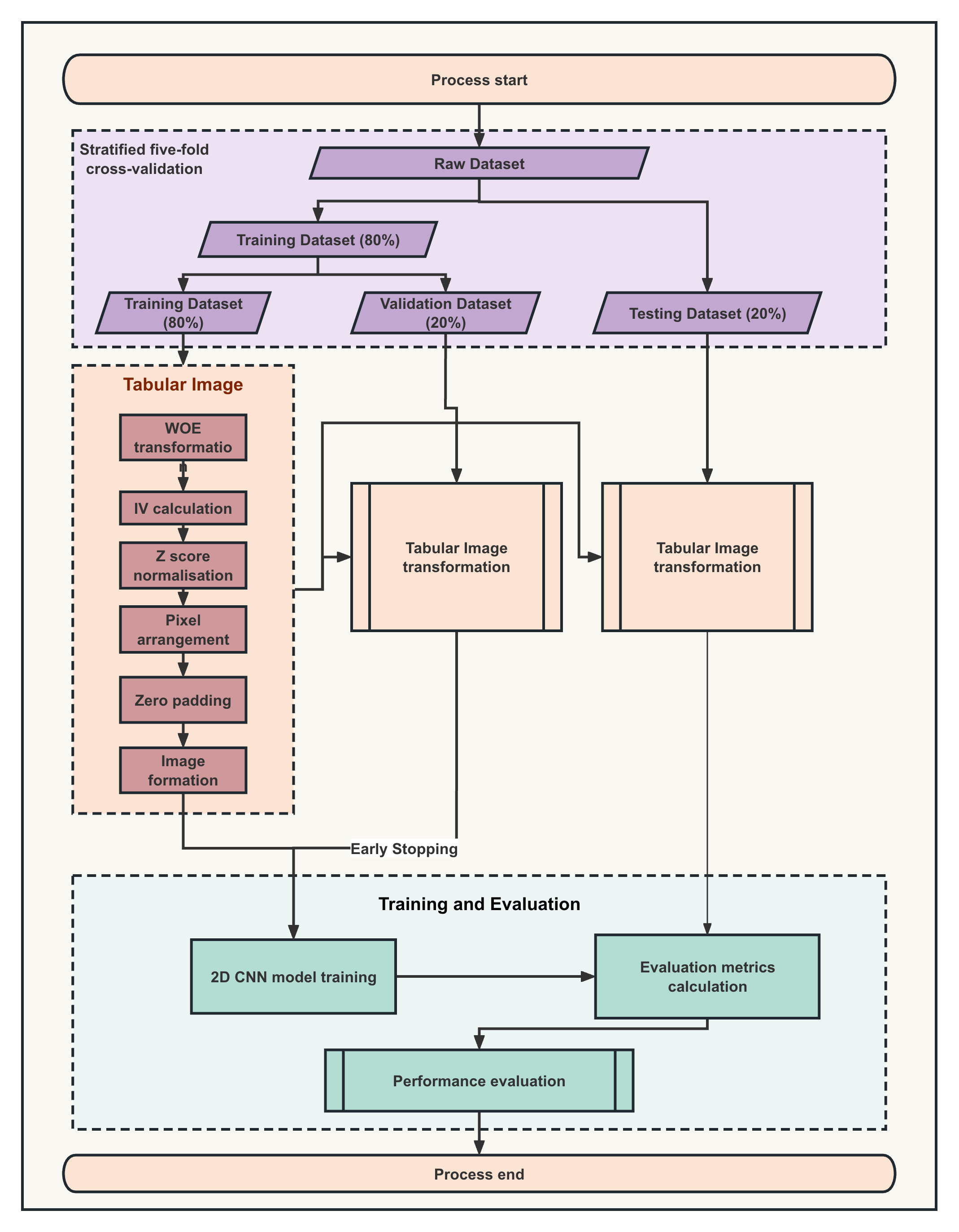}
    \caption{The overall framework of \textit{Tabular Image}}
    \label{fig:flowchartOftabularImage}
\end{figure}

The goal of \textit{Tabular Image} is to transform each sample of tabular data into an image of $N_h \times N_w$ pixels, where $N_h$ and $N_w$ denote the height and the width of the image, respectively. Figure \ref{fig:flowchartOftabularImage} shows the proposed transformation framework. Different from images formed by pixels, tabular data usually contains a mix of numerical and categorical features. To convert tabular data into images, a measure is needed to convert the values of the categorical features to numerical values. Also, the measure needs to represent a feature's ability to separate default and non-default credit applicants, as this is the goal of building credit scoring models. Based on these two requirements, we selected a classic data preprocessing method in credit scoring, the WOE transformation \citep{siddiqi_credit_2012}, in which the difference between the proportion of default and non-default credit applicants for each bin replaces the values of the categorical features. 

After transforming categorical features into WOEs, the dataset was normalised using z-score normalisation and assigned to pre-defined $N_h \times N_w$ images based on feature correlation and IVs. All 2D CNNs were trained and tested in this study using five-fold cross-validation for a fair comparison. Note that to avoid misleading performance caused by data leakage \citep{kaufmanLeakageDataMining2012}, the WOEs and IVs were calculated based on the training set only and passed to the test set for the transformation in every iteration in the five-fold cross-validation. In other words, the default/non-default class information in the test set is not used in tabular data-image transformation in the training. 

\subsection{Weight of evidence transformation}
\label{subsec:WOE}
The first step of transforming original tabular data into images is to convert the values in categorical features into WOEs. The same categorical value will be given the same WOE. The WOE of categorical value $i$ can be defined as:

\begin{equation}\label{eq:woe}
WOE_{i} = \ln{(\frac{N_{bi}}{N_B})} - \ln{(\frac{N_{gi}}{N_G})} 
\end{equation}
where $N_{bi}$ denotes the number of default ('bad') credit applicants with categorical value $i$, $N_{gi}$ denotes the number of non-default ('good') credit applicants with categorical value $i$, $N_B$ denotes the total number of default credit applicants in the (training) dataset, and $N_G$ denotes the total number of non-default credit applicants in the (training) dataset. A large WOE means a strong relationship exists between the categorical value and the binary target variable in identifying default credit applicants.

\subsection{Information value}
\label{subsec:IV}
When the calculation of WOEs is completed, the information value of each feature is calculated to evaluate the feature importance in separating default and non-default credit applicants. IV of feature $i$ can be defined as:
\begin{equation}
    IV = \sum_{i=1}^{n}(\frac{N_{bi}}{N_B} - \frac{N_{gi}}{N_G})*WOE_{i}
\end{equation}
where $n$ denotes the number of categorical values in each feature. For numerical features, values are first discretised into $10$ bins using a quantile-based discretisation \citep{thomas_credit_2017} before calculating the WOEs of each feature. Note that WOEs of numerical features are only used to calculate IVs of numerical features.

\subsection{Feature arrangement}
\label{subsec:featureArr}
\begin{figure}[h!]
    \centering
    \includegraphics[width=\textwidth]{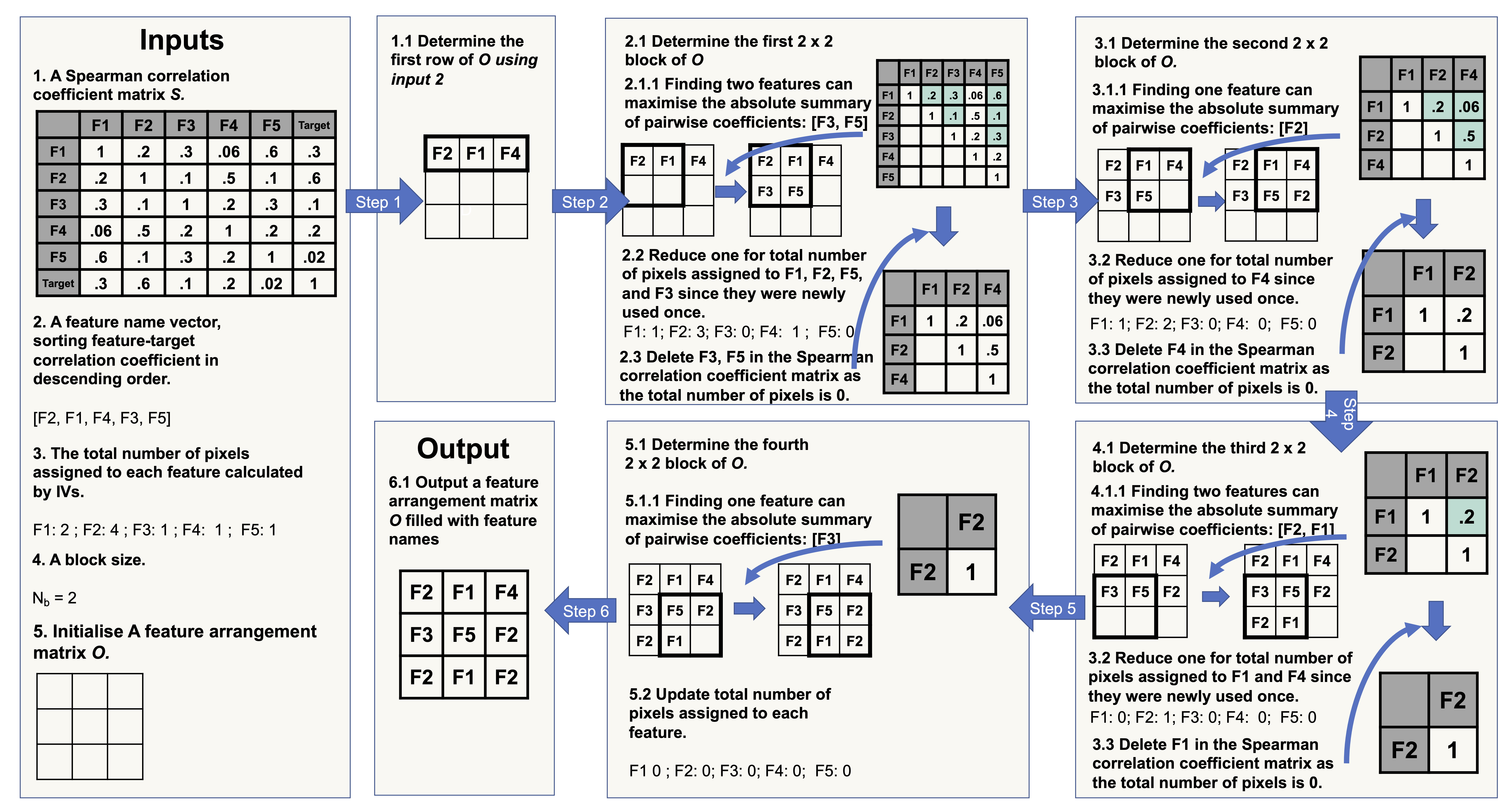}
    \caption{\textbf{Example}: A simple illustration of the feature arrangement process of \textit{Tabular Image}.}
    \label{fig:imageFeatureArrangement}
\end{figure}

In the context of CNNs, the pixels in an image are spatially related, meaning pixels next to each other represent relevant information or patterns. However, in tabular data, such spatial relationships between features don't always exist. Thus, the problem in transforming \textit{Tabular Image} is to allocate pixels in adjacent areas to represent feature values in tabular data while the spatial locations of pixels are still meaningful for tabular data features. In this study, we aim to maximise the Spearman correlation among features in a certain block of the image to represent the spatial relationship as Spearman correlation can assess both linear and non-linear correlations without assuming the frequency distribution of the input variables \citep{haukeComparisonValuesPearson2011}. Figure \ref{fig:imageFeatureArrangement} provides an example illustrating the feature arrangement process. The pseudocode of the feature arrangement method is shown in algorithm \ref{alg:featureArrangement}.

To achieve the goal, firstly, two input parameters, the height $N_{h}$ and the width $N_{w}$ of a tabular image, are set up to define the image size. The product of these two parameters defines the total number of pixels $N_P$ in a tabular image that can be assigned to features in the transformation. The transformation then calculates the total number of pixels assigned to each feature. We define $N_{pi}$, the total number of pixels assigned to each feature $i$, by finding the proportion of IV of each feature in the sum of IV of all features:
    \begin{equation}\label{eq:pixel_i}
        N_{pi} = floor(\frac{IV_{i}}{IV_{T}} * (N_P - N_f)) + 1
    \end{equation}
where $floor(\cdot)$ is a function that rounds numbers down to the nearest integer. $N_P$ denotes the total number of pixels in a tabular image, $N_f$ denotes the number of features in the dataset, $IV_{i}$ denotes the IV of feature $i$, and $IV_{T}$ denotes the sum of the IV of all features. The reason for using rounding down rather than rounding up is that it can avoid the possibility of the sum of $N_{pi}$ of all features exceeding $N_P$. Note that each feature is assigned to at least one pixel. 

After calculating $N_{pi}$, a block size, $N_b$, is defined, indicating the adjacent area of the image that needs to be maximised. Then, we calculate the Spearman correlation coefficient between features and the target variable (default or non-default class). A feature name vector, $V$, is created by sorting the Spearman correlation coefficient between features and the target variable in descending order. After creating $V$, we calculate the Spearman correlation coefficient matrix $S$ among features. Once the Spearman correlation coefficient matrix is calculated, we start to maximise the absolute value of the Spearman correlation coefficient in a block of size $N_b \times N_b$ (see line \ref{alg:featureArrangement:TransStart} to line \ref{alg:featureArrangement:TransEnd} in Algorithm \ref{alg:featureArrangement}). The maximisation process can be seen and solved as a standard 0/1 integer programming problem\footnote{Approximate optimal solutions can be found by using Python package \textit{pyomo} with \textit{gurobi} solver} \citep{hanConvolutionalNeuralNetwork2019}. To begin the maximisation process, we assign the first $N_{w}$ feature in $V$ to the first row of the feature arrangement matrix $O$ to reduce the computation complexity of the optimisation process \citep{hanConvolutionalNeuralNetwork2019}. Then, we iteratively find the feature names that should be used in each block from the upper left to the bottom right of the feature arrangement matrix.

\begin{figure}[h!]
  \centering
\begin{minipage}{\linewidth} 
\begin{algorithm}[H]
\scriptsize
\caption{Feature arrangement method}\label{alg:featureArrangement}
\begin{algorithmic}[1]
\State Input: Spearman correlation coefficient matrix $S$; 
\State Input: Feature name vector $V$, sorting by feature to target feature Spearman correlation coefficient in descending order; 
\State Input: The total number of pixels assigned to each feature $N_{pi}$; 
\State Input: block size $N_b$;
\State Initialise a block $B$ with block size $N_b \times N_b$
\State Initialise total pixel number $N_P$ = $N_{h}$*$N_{w}$ ;
\State Initialise Feature arrangement matrix $O = \varnothing$;
\State Determine the first row of $O$: $O_{1,:} = V_{1:N_{w}}$
\For{each block $B$ in $O$}:\label{alg:featureArrangement:TransStart}
    \State Find known feature names vector and their corresponding Spearman correlation coefficient in the current block;
    \State Find the number of unknown feature names in the current block;
    \State Apply $S$ to obtain a feature names vector that can maximise the sum of the Spearman correlation coefficient in the current block;
    \State Fill the current block $B$ with the known feature names vector and the newly obtained feature names;
    \State If a feature is newly obtained in $B$, update the total number of pixels assigned to each feature $N_{pi} = N_{pi} - 1$;
    \State If the total number of pixels assigned to a feature $N_{pi} = 0$, delete the corresponding Spearman correlation coefficient in $S$;
    \State If all $N_{pi} = 0$, end the for loop.
\EndFor \label{alg:featureArrangement:TransEnd}
\State Output: a feature arrangement matrix $O$.
\end{algorithmic}
\end{algorithm}
\end{minipage}
\end{figure}

\subsection{Tabular Image}
\label{subsec:tabularImageTrans}
\begin{figure}[h!]
    \centering
    \includegraphics[width=\textwidth]{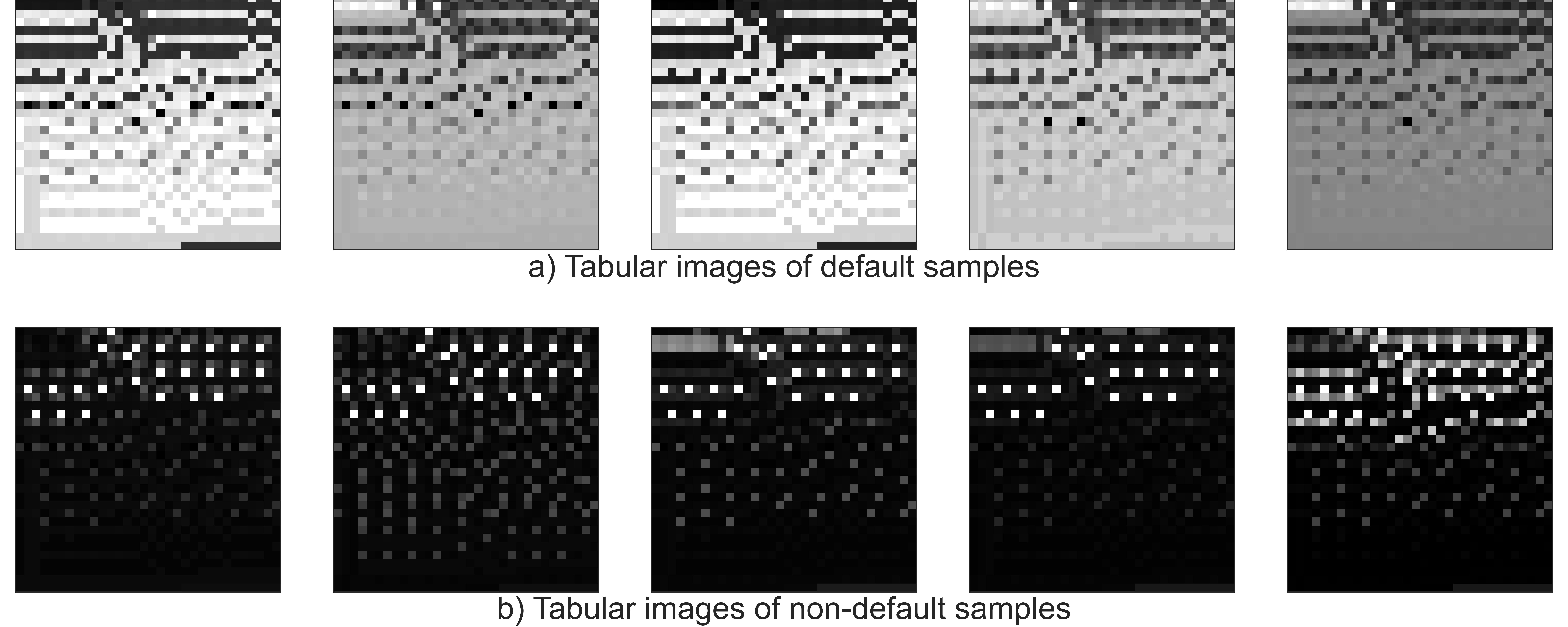}
    \caption{Example images of the TC dataset. (a) is images of five default samples in the TC dataset generated by proposed \textit{Tabular Image}. (b) is images of non-default samples in the TC dataset generated by proposed \textit{Tabular Image}.}
    \label{fig:tabularImageExamples}
\end{figure}

Once WOEs, IVs and a feature arrangement matrix are calculated, a \textit{Tabular Image} transformation can be conducted. The pseudocode of \textit{Tabular Image} is shown in algorithm \ref{alg:tabularImage}. The \textit{Tabular Image} transformation first applies a z-score normalisation to feature values to reach faster convergence. After the normalisation, the \textit{Tabular Image} transformation iteratively converts each tabular data sample into an image. Given a tabular data sample, \textit{Tabular Image} transformation iteratively creates a pixel matrix for each feature by replacing feature names in the feature arrangement matrix with the corresponding feature value. After creating a pixel matrix for all features, padding is applied to the image using the median value of the pixel matrix if the number of pixels assigned to all features is less than the initial $N_P$. The number of paddings for the image is calculated by 
    \begin{equation}\label{eq:zeros}
        N_{paddings} = N_P - \sum_{i=1}^{n}N_{pi}
    \end{equation}
where $n$ is the total number of features. After padding, an image or a pixel matrix that is transformed from one row (i.e., one sample) of the original tabular data is created. After transforming every sample in the dataset, the tabular images can be used as inputs to 2D CNNs. Figure \ref{fig:tabularImageExamples}a shows five default tabular images and figure \ref{fig:tabularImageExamples}b shows five non-default tabular images.\footnote{These default/non-default images are converted from samples in the later introduced Taiwan Credit dataset that have the highest/lowest five probability of default calculated by our later constructed 2D CNN for comparison purposes} The grey level of a normalised pixel represents the normalised value of the corresponding sample. Figure \ref{fig:tabularImageExamples} shows clearly different visual patterns between the default and non-default credit applicants, where the tabular images of non-default credit applicants are darker than default credit applicants. This also indicates that the transformation of tabular data into images might offer unique opportunities for data visualisation and interpretation, such as the heat maps in financial analysis \citep{argyriouFraudDetectionVisualization2014}, credit risk visualisation tools \citep{leiteEVAVisualAnalytics2018}, etc., aiding in a quicker and more intuitive understanding of the credit data.  

\begin{figure}[h!]
  \centering
\begin{minipage}{\linewidth} 
\begin{algorithm}[H]
\scriptsize
\caption{Tabular Image}\label{alg:tabularImage}
\begin{algorithmic}[1]
\State Input: a dataset with $i$ features and $n$ rows;
\State Discretised the numerical features using \textit{qcut} in Python package \textit{pandas};
\State Transform features into WOEs;
\State Calculate $IV_i$ for each feature $i$;
\State Set the image height, $N_{h}$ and the image width, $N_{w}$;
\State Initialise total pixel number $N_P$ = $N_{h}$*$N_{w}$ ;
\State Calculate the number of pixels $N_{pi}$ assigned to each feature $i$; \label{alg:tabularImage:NumberOfPixels}
\State Calculate the Spearman's correlation between each pair of feature $i$ and the Spearman's correlation between each feature and the target feature; 
\State Set the block size $N_b$;
\State Optimise the feature arrangement matrix $O$ based on the number of pixels $N_{pi}$ assigned to each feature, Spearman's correlation matrix and the block size $N_b$;
\State Initialise $N_{total} = \sum_{i=1}^{n}N_{pi}$;
\State Applies the z-score normalisation to the dataset; \label{alg:tabularImage:minMax}
\For{row $i, i=[1, 2, ..., n]$ in the dataset} \label{alg:tabularImage:rowTransStart}
    \State Initialise pixel matrix $P_i$ = $O$
    \State Replace feature names in $P_i$ with the feature's corresponding numerical value;
    \State Replace Padding in $P_i$ with the median value in the pixel matrix;
\EndFor \label{alg:tabularImage:rowTransEnd}
\State Output: a pixel matrix list $P_1, P_2, ..., P_n$.
\end{algorithmic}
\end{algorithm}
\end{minipage}
\end{figure}

\section{Experiments}
\label{sec:experiments}
\subsection{Data Preparation}
\label{subsec:experiments:data}
To evaluate the performance of \textit{Tabular Image} on 2D CNNs, we applied the transformation on a standard benchmark dataset named Taiwan Credit (TC)\footnote{see: \url{https://archive.ics.uci.edu/dataset/350/default+of+credit+card+clients}} \citep{yeh_comparisons_2009} which is widely used in credit scoring literature \citep{dumitrescu_machine_2022, jiangBenchmarkingStateoftheartImbalanced2023, shiImprovedCreditRisk2024}. Then, a more complex dataset, namely Home Credit Default Risk (HC)\footnote{see: \url{https://www.kaggle.com/c/home-credit-default-risk/overview}} \citep{home-credit-default-risk} was considered. Home Credit is a multinational consumer finance provider focusing on instalment lending primarily to borrowers with little or no credit history, adding to the data's complexity. Finally, another complex dataset, namely Fannie Mae (FM)\footnote{see: \url{https://capitalmarkets.fanniemae.com/credit-risk-transfer/single-family-credit-risk-transfer/fannie-mae-single-family-loan-performance-data}} was selected. Fannie Mae is a government-sponsored enterprise that mainly focuses on mortgage loans in the US. We define a loan as default if a loan is more than 90 days past the due date (DPD)\footnote{We consider the loan as default even if the loan is cured after 90 DPD} based on major financial standards such as Basel II and IFRS 9 \citep{mushavaNovelXGBoostExtension2022}. The original dataset contains more than ten million loan records. Due to the computing resource limitation, stratified sampling was performed to randomly sample data from a relatively stable economic period from 2009 to 2016 \citep{chenPredictingMortgageEarly2021}. Features that contain $99\%$ or more missing values are dropped. The details of the three datasets are shown in Table \ref{table:datasetsInfo}. All the datasets are imbalanced, which matches real-world situations in a credit scoring context, as defaulting credit applicants are far fewer than non-defaulting ones. The HC and FM datasets present significant challenges, particularly due to their large sample sizes and notably lower default rates. The HC dataset, in particular, poses additional complexities with its large feature sizes and samples primarily belonging to the unbanked population.

The complexity and real-world applicability of these datasets make them a rigorous testing ground for different models. In this context, even a modest improvement in performance is significant and may lead to huge economic benefits \citep{blochlingerEconomicBenefitPowerful2006}.

\begin{table}[hbt!]
\caption{Information of credit datasets used in this study}
\begin{tabular}{lllll}
\hline
Dataset & Sample size & No. of features & Default samples & Default rate \\ \hline
TC & 30000 & 23 & 6636 & 22.1\%\\
HC& 307511& 120& 24825& 8.1\%\\
FM& 300000& 36& 925& 0.3\%\\ \hline
\end{tabular}%
\label{table:datasetsInfo}
\end{table}

We separately applied the \textit{Tabular Image} transformation on TC, HC and FM datasets to generate their tabular images. The size of each image was set to $32\times32=1024$ pixels with a block size $3\times3$ pixels\footnote{The block size in our model is tuned as a hyper-parameter. A block size of 3 × 3 is optimal for our study}.

A standard pre-processing method, as detailed by \citet{gunnarsson_deep_2021}, was applied when using tabular data as input for logistic regression, support vector machine (SVM), decision tree (DT), random forest, Adaptive Boosting (AdaBoost), Gradient-boosted decision trees (GBDT),  XGBoost, MLP, and 1D CNN. First, categorical features were transformed into WOEs, with missing values being regarded as separate categorical values. Then, missing values in numerical features were imputed with their median values. After processing missing values, random oversampling was applied to the training set, a technique proven effective in addressing data imbalance issues in the credit scoring field \citep{jiangBenchmarkingStateoftheartImbalanced2023}. After that, features were standardised using z-score normalisation. 

For comparison purposes, we also transformed the TC, HC and FM datasets into images using the One-hot \citep{dastile_making_2021} and DeepInsight \citep{sharma_deepinsight_2019} transformation methods. For One-hot transformation, we first discretised numerical features using the quantile-based discretisation method, akin to the \textit{Tabular Image} approach. Then, sparse binary pixel matrices \citep{dastile_making_2021}, which are matrices consisting of values $0$ and $1$, were created to represent the one-hot encoding of each feature. Features with IV larger than $0.1$ were selected, as suggested by \citet{dastile_making_2021}. The image size was resized to $32\times32$ as the original image size does not meet the minimum input requirement of the 2D CNN used. For the DeepInsight transformation, we first pre-process tabular data using the method suggested by \citet{gunnarsson_deep_2021}. After this, tabular data was transformed into images with a size of $32\times32$ pixels.

\subsection{Models Construction}
\label{subsec:experiments:models}

This study constructed ten prediction models\footnote{For Logistic regression, SVM, DT, Random forest, AdaBoost, GBDT and MLP, we used the implementation from a Python machine learning library called Scikit-Learn (see: \url{https://scikit-learn.org/}). For XGBoost, we used the XGBoost library (see: \url{https://xgboost.ai/}) in Python. For 1D and 2D CNNs, models were constructed using Pytorch (see: \url{https://pytorch.org/}).}, including logistic regression \citep{wiginton_note_1980}, SVM \citep{cortesSupportvectorNetworks1995}, DT \citep{breimanClassificationRegressionTrees1984}, random forest \citep{brown_experimental_2012,wang_two_2012}, AdaBoost \citep{freundDecisionTheoreticGeneralizationLine1997}, GBDT \citep{friedmanGreedyFunctionApproximation2001}, XGBoost \citep{lessmann_benchmarking_2015}, MLP \citep{gunnarsson_deep_2021}, 1D CNN \citep{shwartz-ziv_tabular_2022}, and a 2D CNN named ConvNeXt \citep{wooConvNeXtV2Codesigning2023}. Logistic regression was selected as an industry standard. SVM and DT were chosen as benchmark individual classifiers. Random forest was chosen as a benchmark bagging ensemble classifier as suggested by \citet{lessmann_benchmarking_2015}, along with two mainstream boosting ensemble models, AdaBoost and GBDT \citep{shiImprovedCreditRisk2024}. XGBoost was regarded as the state-of-the-art ensemble classifier in this study as it yields superior performance among machine learning models \citep{gunnarsson_deep_2021, grinsztajnWhyTreebasedModels2022a}. As 2D CNNs are not suitable for tabular data \citep{damriEfficientImagebasedRepresentation2023}, we utilised a 1D CNN and an MLP to analyse the performance of neural networks on tabular data. ConvNeXt was selected to train our proposed \textit{Tabular Image} because it is a state-of-the-art 2D CNN architecture and serves as a backbone of advanced 2D CNNs \citep{wooConvNeXtV2Codesigning2023}. Details and hyper-parameter setting of each model are presented in the following paragraphs. 

The ConvNeXt architecture is proposed by \citet{wooConvNeXtV2Codesigning2023}, which is a state-of-the-art 2D CNN architecture upgraded from ResNet \citep{heDeepResidualLearning2016}. By introducing shortcut connections \citep{heDeepResidualLearning2016}, ResNet-like architecture mitigates the vanishing gradient problem when the depth of CNN increases to make the construction of very deep CNN possible and performs remarkably well. Furthermore, the depthwise convolution and global response normalisation (GRN) layers in ConvNeXt allow it to pay attention to specific areas of an image, which works similarly to feature selection, an important step in credit scoring.

We used the Nesterov stochastic gradient descent as the optimiser with a momentum of $0.9$. Binary cross entropy was selected as the loss function. The batch size was set to 128 for the TC dataset and 1024 for the HC dataset. The learning rate was initialised as $0.001$ for TC and HC and $0.0001$ for the FM dataset. The initial learning rate was divided by ten every ten epochs during training to avoid overfitting. Early stopping was also used to avoid overfitting by stopping the training process when the AUC of the validation set stopped increasing. A patient of five was set to avoid local minima.

The architecture of the 1D CNN is similar to LeNet-5 \citep{lecun_backpropagation_1989}, comprising one input layer, two 1D convolutional layers with ReLU activation function, two average pooling layers, two fully connected layers and one output layer. The batch size, optimiser, and loss function were set identically to ConvNeXt. The learning rate was initialised as $0.1$ and divided by ten every ten epochs. 

Also, a 5-layer MLP was constructed to evaluate the effect of fully connected artificial neural networks with tabular data. The ReLU activation function was used in each hidden layer. The optimiser and loss function were set identically to ConvNeXt. The batch size was set to 200 for all datasets.

\subsection{Hyper-parameter tunning}
\label{subsec:experiments:hypertune}
To guarantee a fair and rigorous comparison among the various machine learning models, including ours, we conducted an extensive hyper-parameter tuning process for each model to achieve their optimal performance. This involved a meticulous grid search to identify the best hyper-parameters for each model. The hyper-parameter search space is shown in Table \ref{tab:hyperpara}. The chosen values of each model were either recommended by literature (e.g. \citet{gunnarsson_deep_2021}) or derived from our own exploratory analysis. This thorough approach to tuning ensures that the performance results presented are the best possible for each model. Consequently, even slight improvements achieved by our model carry significant weight, underlining its effectiveness compared to others.

\begin{table}[hbt]
\centering
\caption{Hyper-parameter tuning grid}
\begin{tabular*}{\textwidth}{lp{2.5cm}p{3cm}p{3cm}}
\hline
 & Number of Models need to be search & Hyper-parameters & Grid range \\ \hline
Logistic regression & 1 & - & - \\
 &  &  &  \\
SVM & 6 & C & 0.1, 1, 10 \\
 &  & Gamma & $1/(n*Var(X)), 1/n$ \\
 &  &  &  \\
DT & 54 & Minimum number of samples for each split & 2, 3, 4, 5, 6, 7, 8, 9, 10 \\
 &  & Minimal cost-complexity pruning & 0, 0.1, 0.2, 0.3, 0.4, 0.5 \\
 &  &  &  \\
Random forests & 30 & Number of trees & 100, 250, 500, 750, 1000 \\
 &  & Number of features & $\sqrt{n}$, $\log_2(n)$ \\
 &  & Ratio of samples to be selected & 0.5, 0.75, 1 \\
 &  &  &  \\
AdaBoost & 24 & Number of gradient boosted trees & 50, 100, 150, 200, 300, 500 \\
 &  & Learning rate & 0.1, 0.2, 0.3, 0.4 \\
 &  &  &  \\
GBDT & 216 & Number of gradient boosted trees & 50, 100, 150, 200, 300, 500 \\
 &  & Maximum tree depth & 1, 2, 3 \\
 &  & Learning rate & 0.1, 0.2, 0.3, 0.4 \\
 &  & Ratio of samples to be selected & 0.5, 0.75, 1 \\
 &  &  &  \\
XGBoost & 216 & Number of gradient boosted trees & 50, 100, 150, 200, 300, 500 \\
 &  & Maximum tree depth & 1, 2, 3 \\
 &  & Learning rate & 0.1, 0.2, 0.3, 0.4 \\
 &  & Ratio of samples to be selected & 0.5, 0.75, 1 \\
 &  &  &  \\
MLP & 253 & Number of hidden units in each layer & 5, 10, 15, 20, 25, 40, 100 \\
 &  & Number of layers & 1, 5 \\
 &  & Learning rate & 0.001, 0.0001, 0.00001 \\
 &  & Strength of the L2 regularization & 0, 0.001. 0.01, 0.1 \\
 &  &  &  \\
Tabular Image & 3 & Block size & 2, 3, 4
 \\ \hline
\end{tabular*}%
\footnotetext[1]{MLPs that have a growing number of hidden units in layers were not taken into account as they tend not to generalise well.}
\label{tab:hyperpara}
\end{table}

\subsection{Evaluation metrics}
\label{subsec:experiments:metrics}

This study evaluates a model's overall performance and the maximum ability to separate default and non-default credit applicants using three metrics that are commonly used in credit scoring \citep{baesens_benchmarking_2003, lessmann_benchmarking_2015}, the area under the receiver operating characteristic curve (AUC), the Kolmogorov-Smirnov statistic (KS) and the H-measure. The AUC is a metric that evaluates the overall discrimination ability of a model by measuring the area under the receiver operating characteristic (ROC) curve. In credit scoring tasks, it is equivalent to the probability that the credit score of a randomly chosen default credit applicant is higher than a randomly chosen non-default credit applicant \citep{lessmann_benchmarking_2015}. Since the AUC considers a model's global performance, it assumes that all thresholds are equally possible to use as cutoff points in credit scoring, which is not plausible in practice \citep{hand_good_2005}. Because of this, it is also essential to find the cutoff point that can maximise the distance between default and non-default credit applicants in order to evaluate the maximum ability of a model to separate default and non-default credit applicants. Therefore, the KS was selected to concentrate on measuring the maximum distance between default and non-default credit applicants that a model can separate (i.e. the KS point). The H-measure \citep{handMeasuringClassifierPerformance2009} is a coherent alternative metric compared to the AUC. It is equivalent to the percentage improvement of the expected minimum loss a classifier gains compared to a classifier that randomly assigns samples to classes. H-measure allows us to specify a misclassification cost during evaluation, which is essential since misclassifying a default borrower to non-default is considered a more severe case in credit scoring applications. In this study, the misclassification cost was set to the number of default samples divided by the number of non-default samples \citep{handBetterBetaMeasure2014}. Lastly, these three metrics, the AUC, the KS and the H-measure, are robust toward data imbalance \citep{lessmann_benchmarking_2015}, which is vital since the datasets used in this study show various degrees of data imbalance.

\section{Results}
\label{sec:res}

\subsection{Distribution analysis of tabular data and tabular images}
\label{subsec:results:distributionanalysis}

\begin{figure}[hbt!]
\centering
\includegraphics[width=0.9\textwidth]{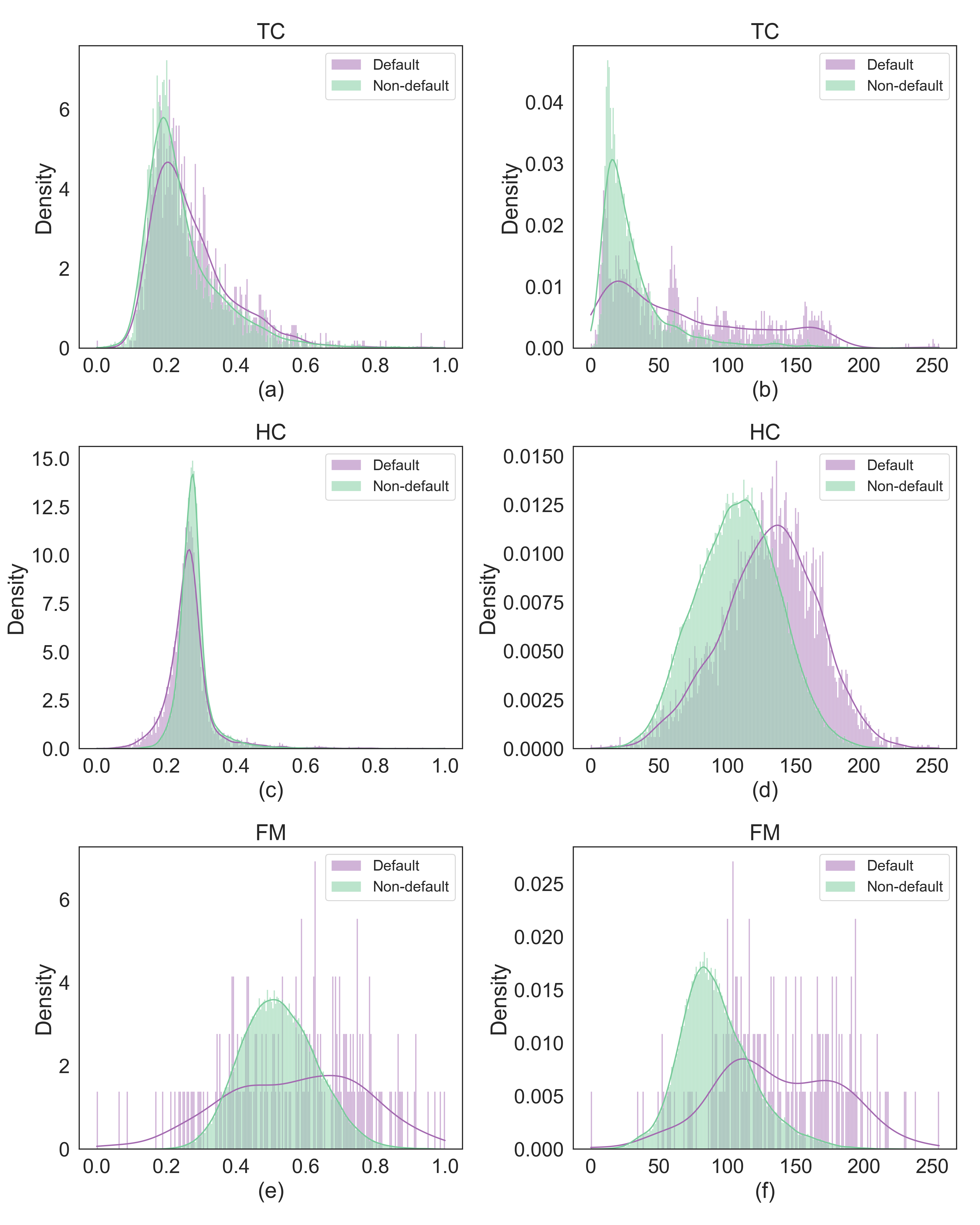}
\caption{The density plot of mean pixel values of tabular images and the density plot of mean tabular data value for the test set in the TC, HC, and FM datasets. The range of the X-axis in Figure \ref{fig:mean_pixel}b, Figure \ref{fig:mean_pixel}d, and Figure \ref{fig:mean_pixel}f is from 0 (black pixels) to 255 (white pixels), which indicates the pixel value range of grayscale images. The range of the X-axis in Figure \ref{fig:mean_pixel}a, \ref{fig:mean_pixel}c, \ref{fig:mean_pixel}e is from 0 to 1, which indicates the tabular data value after min-max normalisation.}
\label{fig:mean_pixel}
\end{figure}

To investigate the effectiveness of \textit{Tabular Image}, we plotted three mean pixel value density plots and three mean tabular data value density plots by using tabular images and tabular data converted from samples in the test set of the TC, HC, and FM datasets. Figure \ref{fig:mean_pixel}b, \ref{fig:mean_pixel}d, and \ref{fig:mean_pixel}f are three mean pixel value density plots of the TC, HC, and FM datasets, respectively. The range of the X-axis in Figure \ref{fig:mean_pixel}b, \ref{fig:mean_pixel}d, and \ref{fig:mean_pixel}f is from 0 (black pixels) to 255 (white pixels), which indicates the pixel value range of grayscale images.  Figure \ref{fig:mean_pixel}a, \ref{fig:mean_pixel}c, and \ref{fig:mean_pixel}e are three mean tabular data value density plots of the TC, HC, and FM datasets, respectively. The range of the X-axis in Figure \ref{fig:mean_pixel}a, \ref{fig:mean_pixel}c, and \ref{fig:mean_pixel}e is from 0 to 1, which indicates the tabular data value after min-max normalisation. The distribution of non-default and default samples is significantly overlapped in Figure \ref{fig:mean_pixel}a, \ref{fig:mean_pixel}c, and \ref{fig:mean_pixel}e, indicating the difficulty in distinguishing these two distributions using raw tabular data. However, it is apparent that after being transformed by \textit{Tabular Image}, the separation of the distribution of non-default and default samples is more pronounced. Specifically, in the transformed tabular images, the majority of pixels in non-default tabular images are distributed in the left part of the X-axis, which means they are generally darker than those in the default tabular images, which aligns with the observation of Figure \ref{fig:tabularImageExamples} in section \ref{subsec:tabularImageTrans}. The more pronounced separation between the distribution of non-default and default samples further demonstrated the power of \textit{Tabular Image}, indicating the effectiveness of our method.

\subsection{Comparison of 2D CNN with machine learning models}
\label{subsec:results:comparewithml}

To better evaluate the discriminatory performance of the \textit{Tabular Image}, we implemented a stratified five-fold cross-validation process to ensure the reliability of the results. In each iteration, one fold was reserved as the test set, which had not been previously encountered by the model, thereby evaluating the performance of the model. For the other four folds, we used $80\%$ of the samples as the training set and the remaining $20\%$ of the samples as the validation set for hyper-parameter tuning. We first evaluated traditional prediction models with original tabular data, including XGBoost, Random forest, Logistic regression, and MLP, as a baseline. Those baseline models tell us how good the performance can be with the original tabular data. Then we used our proposed \textit{Tabular Image} with a ConvNeXt. Because ConvNeXt cannot accept tabular data as input, we also ran a 1D CNN on the original tabular data to examine the effect of the CNN when using tabular data. In order to discuss the effects of different tabular data-image transformation methods on prediction performance, we also applied the One-hot transformation \citep{dastile_making_2021} and DeepInsight \citep{sharma_deepinsight_2019} on ConvNeXt, respectively. Moreover, we also ran the \textit{Tabular Image} with different image sizes and allocation methods to assess the robustness of our proposed method.

\subsubsection{Performance of 2D CNN and machine learning models}
\label{subsec:results:performanceofml}
We first compared 2D CNN using tabular images with traditional prediction models. Tables \ref{table:mlmethodsresTC1}, \ref{table:mlmethodsresHC1} and \ref{table:mlmethodsresFM1} present the average test set AUC, H-measure, KS, and their corresponding standard deviations from five-fold cross-validation for each prediction model on the TC, HC and FM datasets, respectively.

On the smaller TC dataset, with the proposed \textit{Tabular Image}, a 2D CNN outperformed all individual classifiers, including its 1D counterpart, 1D CNN, by approximately $1\%$ to $20\%$ in terms of AUC and H-measure. Note that 1D CNN performed only slightly better than the weakest logistic regression in terms of AUC and H-measure. These results indicated that a 2D ConvNeXt could utilise 2D kernels to better capture features in the data after transforming tabular data into images than 1D CNN that extracts signals in 1D tabular data with 1D kernels. Meanwhile, as an individual classifier, our method outperformed the ensemble benchmark classifier, random forest, on all evaluation metrics and achieved similar results compared to the state-of-the-art GBDT and XGBoost. These results indicate that our method captured linear and non-linear relationships in the data better after transforming tabular data into images than neural networks directly using tabular data as input. Our results also confirmed the findings in \citet{gunnarsson_deep_2021} that XGBoost can outperform other benchmark models with traditional tabular data on small datasets.

On the HC dataset, which has a much larger sample size and more challenging testing conditions, the ConvNeXt with tabular images outperformed individual classifiers by approximately $1\%$ to $22\%$ in terms of AUC and H-measure. Notably, our method even outperformed the ensemble classifiers, including the state-of-the-art GBDT and XGBoost, across all evaluation metrics. It becomes even more pronounced regarding the H-measure, particularly with an increment of $1.39\%$, whereas other benchmark models suffered a decrease ranging from $0.5\%$ to $21.1\%$ compared to GBDT and XGBoost. 

The FM dataset, which is large and extremely imbalanced, presented a different challenge. Again, the ConvNeXt with tabular images outperformed individual classifiers with an increase of approximately $5\%$ to $29\%$ in terms of AUC and H-measure. What stands out in this table is that our method outperformed the GBDT and XGBoost across all evaluation metrics. The improvement becomes even more pronounced regarding the H-measure with an improvement of $4.37\%$ from GBDT and $5.68\%$ from XGBoost. 

To further investigate the predictive performance of our method, we particularly analyse subprime borrowers\footnote{A rule of thumb is that a subprime borrower is one who has a FICO score lower than 670. See https://www.experian.com/blogs/ask-experian/what-is-subprime/} in the FM dataset. Subprime borrowers are considered riskier for a lender and difficult to predict. Thus, a model that can perform well on subprime borrowers can be considered more advanced and help reduce loan loss. Similar to the previous results format, Table \ref{table:mlmethodsresFMlow1} shows the performance of each prediction model on subprime samples. We can see that ConvNeXt with tabular images outperformed all individual and ensemble classifiers in terms of AUC, H-measure, and KS with an improvement of approximately $8\%$ to $21\%$, $22\%$ to $46\%$, $0.20$ to $0.35$, respectively.

The success of our model in those datasets highlights the effectiveness of our data transformation method, indicating a higher level of robustness. On the one hand, our method adapts better to varying data complexities and sizes, which is crucial in practical applications \citep{grinsztajnWhyTreebasedModels2022a}. On the other hand, with the \textit{Tabular Image}, ConvNeXt can better extract both linear and non-linear relationships in a complicated dataset, making it more powerful in prediction, demonstrating the effectiveness of the proposed \textit{Tabular Image}. Interestingly, the performance of 1D CNN with tabular data is worse than that of logistic regression in the HC and FM datasets, with a decrease of more than $3\%$ across all three metrics, indicating 1D CNN does not have sufficient ability to extract information in such large-scale, challenging datasets. In contrast, by converting tabular data into images, we unlock the full potential of the 2D CNN and identify an improvement of approximately$1\%-5\%$ across all three metrics compared to the classic logistic regression model. This further emphasises the transformative impact of our \textit{Tabular Image}, showcasing its unique contribution to advancing the capabilities of 2D CNNs in credit scoring.

Although we mainly focus on predictive performance, we noticed that the training time of ConvNeXt and the conversion time of \textit{Tabular Image} are relatively short. It took less than ten epochs for a ConvNeXt trained on the TC dataset to converge, with a training time of less than two minutes on two P100 GPUs. The model trained on the HC and the FM datasets took about ten epochs to converge, with a training time of less than $15$ minutes on two P100 GPUs. As a complicated dataset with more noise and samples, the increase in training time in the HC and the FM datasets was expected, but one can easily reduce the training time by using multiple GPUs or a more advanced one\footnote{We observed a reduction of $87\%$ of the training time to less than two minutes as we tested on one Nvidia L40s GPU compared to training on two P100 GPUs.}. The \textit{Tabular Image} involves a feature arrangement step, as described in Algorithm 1, and a tabular data-to-image conversion step, as described in Algorithm 2. Similar to tabular data transformation techniques, such as normalisation in the preprocessing step, the feature arrangement step is a one-time calculation in order to generate a feature arrangement matrix. The feature arrangement matrix can be saved and used directly in the transformation of incoming data to reduce the computational cost. For the tabular data-to-images conversion step, the time complexity is linear $O(n)$ since there is only a single loop within the tabular data-to-image conversion step. In our empirical evaluations, the practical runtime of the feature arrangement step was less than three minutes for all the datasets considered in this study, with the use of an Apple M1 CPU. The total runtime of the tabular data-to-images conversion step was less than 15 seconds for the small TC dataset and less than four minutes for the larger HC and FM datasets, using the same CPU. This is a reasonable and manageable trade-off considering the significant improvement in predictive performance. Moreover, it is feasible to implement and further reduce the computational cost in real-world applications with the use of advanced CPUs. The training time of the ConvNeXt model is also reasonable (e.g., minutes rather than seconds compared to traditional XGBoost), especially considering the potential to take advantage of the latest developments in both hardware and software that allow for faster training processes fully optimised and supported for 2D CNNs. Such scalability and efficiency of our proposed method indicate the potential for real-world application at scale \citep{xiaNovelTreebasedDynamic2020}. Because of the thriving online loan applications, the number of loan applications has increased exponentially compared to traditional offline lending. Thus, more samples can be used to train credit scoring models. It is important to consider a more effective approach to using large datasets in order to keep up with the changing trends in lending services. Therefore, the slight and manageable increase in computational cost is outweighed by the enhanced predictive performance of the proposed \textit{Tabular Image} and its subsequent advantages.

\begin{table*}[hbt!]
\centering
\caption{The average (standard deviation) test set AUC, H-measure, and KS in five-fold cross-validation for each prediction model and the 2D CNN with tabular images: TC dataset.}
\label{table:mlmethodsresTC1}
\resizebox{\textwidth}{!}{%
\begin{tabular}{lllll}
\hline
 & Model & AUC & H-measure & KS \\ \hline
Individual   classifier & Logistic Regression & 76.51\% $\pm$ (0.83\%) & 27.61\% $\pm$ (2.06\%) & 0.4160 $\pm$ (0.0224) \\
 & SVM & 76.37\% $\pm$ (1.14\%) & 26.99\% $\pm$ (2.18\%) & 0.4154 $\pm$ (0.0209) \\
 & DT & 63.09\% $\pm$ (0.61\%) & 8.95\% $\pm$ (0.71\%) & 0.2452 $\pm$ (0.0115) \\
 & MLP & 77.12\% $\pm$ (0.71\%) & 28.01\% $\pm$ (1.54\%) & 0.4180 $\pm$ (0.0176) \\
 & 1D CNN & 76.80\% $\pm$ (0.83\%) & 27.85\% $\pm$ (1.67\%) & 0.4214 $\pm$ (0.0171) \\
Ensemble classifier & Random Forest & 77.14\% $\pm$ (0.88\%) & 28.02\% $\pm$ (1.87\%) & 0.4177 $\pm$ (0.0168) \\
 & AdaBoost & 77.50\% $\pm$ (0.92\%) & 28.25\% $\pm$ (1.98\%) & 0.4204 $\pm$ (0.0201) \\
 & GBDT & \textbf{78.00\% $\pm$ (0.92\%)} & 29.09\% $\pm$ (1.90\%) & \textbf{0.4319 $\pm$ (0.0206)} \\
 & XGBoost & 77.99\% $\pm$ (0.89\%) & \textbf{29.14\% $\pm$ (1.84\%)} & 0.4307 $\pm$ (0.0206) \\
Proposed method & ConvNeXt & 77.98\% $\pm$ (1.03\%) & 28.88\% $\pm$ (2.08\%) & 0.4281 $\pm$ (0.0236) \\ \hline
\end{tabular}%
}
\end{table*}

\begin{table*}[hbt!]
\centering
\caption{The average (standard deviation) test set AUC, H-measure, and KS in five-fold cross-validation for each prediction model and the 2D CNN with tabular images: HC dataset.}
\label{table:mlmethodsresHC1}
\resizebox{\textwidth}{!}{%
\begin{tabular}{cllll}
\hline
\multicolumn{1}{l}{} & Model & AUC & H-measure & KS \\ \hline
Individual   classifier & Logistic Regression & 73.72\% $\pm$ (0.97\%) & 17.65\% $\pm$ (1.32\%) & 0.3544 $\pm$ (0.0147) \\
 & SVM & 57.74\% $\pm$ (2.17\%) & 2.85\% $\pm$ (1.19\%) & 0.1181 $\pm$ (0.0296) \\
 & DT & 53.40\% $\pm$ (0.66\%) & 1.29\% $\pm$ (0.36\%) & 0.0675 $\pm$ (0.0126) \\
 & MLP & 73.96\% $\pm$ (0.31\%) & 17.87\% $\pm$ (0.41\%) & 0.3589 $\pm$ (0.0056) \\
 & 1D CNN & 70.61\% $\pm$ (3.58\%) & 13.66\% $\pm$ (4.83\%) & 0.3056 $\pm$ (0.0530) \\
Ensemble   classifier & Random Forest & 74.06\% $\pm$ (0.22\%) & 18.43\% $\pm$ (0.52\%) & 0.3611 $\pm$ (0.0044) \\
 & AdaBoost & 74.04\% $\pm$ (1.03\%) & 18.12\% $\pm$ (1.61\%) & 0.3573 $\pm$ (0.0158) \\
 & GBDT & 74.57\% $\pm$ (0.85\%) & 18.76\% $\pm$ (1.36\%) & 0.3653 $\pm$ (0.0134) \\
 & XGBoost & 74.28\% $\pm$ (0.98\%) & 18.31\% $\pm$ (1.58\%) & 0.3600 $\pm$ (0.0148) \\
Proposed method & ConvNeXt & \textbf{75.06\% $\pm$ (0.36\%)} & \textbf{19.70\% $\pm$ (0.72\%)} & \textbf{0.3729 $\pm$ (0.0093)} \\ \hline
\end{tabular}
}
\end{table*}

\begin{table*}[hbt!]
\centering
\caption{The average (standard deviation) test set AUC, H-measure, and KS in five-fold cross-validation for each prediction model and the 2D CNN with tabular images: FM dataset.}
\label{table:mlmethodsresFM1}
\resizebox{\textwidth}{!}{%
\begin{tabular}{cllll}
\hline
\multicolumn{1}{l}{} & Model & AUC & H-measure & KS \\ \hline
Individual   classifier & Logistic Regression & 87.15\% $\pm$   (1.01\%) & 43.56\% $\pm$ (2.30\%) & 0.5962 $\pm$ (0.0157) \\
 & SVM & 80.86\% $\pm$ (2.29\%) & 34.59\% $\pm$ (3.25\%) & 0.4945 $\pm$ (0.0338) \\
 & DT & 74.63\% $\pm$ (0.84\%) & 26.17\% $\pm$ (1.59\%) & 0.4926 $\pm$ (0.0168) \\
 & MLP & 85.01\% $\pm$ (0.86\%) & 37.85\% $\pm$ (2.00\%) & 0.5558 $\pm$ (0.0256) \\
 & 1D CNN & 84.45\% $\pm$ (1.70\%) & 36.83\% $\pm$ (3.67\%) & 0.5572 $\pm$ (0.0328) \\
Ensemble   classifier & Random Forest & 89.92\% $\pm$ (0.56\%) & 49.44\% $\pm$ (1.61\%) & 0.6439 $\pm$ (0.0086) \\
 & AdaBoost & 88.59\% $\pm$ (1.63\%) & 46.74\% $\pm$ (2.65\%) & 0.6321 $\pm$ (0.0340) \\
 & GBDT & 90.12\% $\pm$ (2.16\%) & 50.68\% $\pm$ (4.03\%) & 0.6446 $\pm$ (0.0483) \\
 & XGBoost & 89.43\% $\pm$ (3.87\%) & 49.19\% $\pm$ (6.43\%) & 0.6434 $\pm$ (0.0787) \\
Proposed method & ConvNeXt & \textbf{91.75\% $\pm$ (1.15\%)} & \textbf{55.05\% $\pm$ (3.21\%)} & \textbf{0.6827 $\pm$ (0.0355)} \\ \hline
\end{tabular}
}
\end{table*}

\begin{table*}[hbt!]
\centering
\caption{The average (standard deviation) test set AUC, H-measure, and KS in five-fold cross-validation for each prediction model and the 2D CNN with tabular images: Subprime samples in the FM dataset.}
\label{table:mlmethodsresFMlow1}
\resizebox{\textwidth}{!}{%
\begin{tabular}{lllll}
\hline
 & Model & AUC & H measure & KS \\ \hline
Individual   classifier & Logistic Regression & 79.21\% $\pm$ (4.81\%) & 29.13\% $\pm$ (7.04\%) & 0.4937 $\pm$ (0.0662) \\
 & SVM & 76.50\% $\pm$ (5.30\%) & 28.01\% $\pm$ (9.64\%) & 0.4401 $\pm$ (0.0689) \\
 & DT & 69.82\% $\pm$ (4.55\%) & 14.15\% $\pm$ (5.53\%) & 0.3963 $\pm$ (0.0910) \\
 & MLP & 76.10\% $\pm$ (4.70\%) & 24.81\% $\pm$ (4.79\%) & 0.4481 $\pm$ (0.0670) \\
 & 1D CNN & 80.34\% $\pm$ (3.65\%) & 31.87\% $\pm$ (4.94\%) & 0.5139 $\pm$ (0.0596) \\
Ensemble classifier & Random Forest & 84.53\% $\pm$ (2.17\%) & 37.71\% $\pm$ (5.16\%) & 0.5498 $\pm$ (0.0485) \\
 & AdaBoost & 80.84\% $\pm$ (1.79\%) & 30.98\% $\pm$ (2.39\%) & 0.5213 $\pm$ (0.0473) \\
 & GBDT & 83.13\% $\pm$ (4.87\%) & 38.06\% $\pm$ (8.17\%) & 0.5329 $\pm$ (0.0753) \\
 & XGBoost & 83.45\% $\pm$ (2.59\%) & 35.95\% $\pm$ (4.01\%) & 0.5139 $\pm$ (0.0596) \\
Proposed method & ConvNeXt & \textbf{91.06\% $\pm$ (5.20\%)} & \textbf{60.32\% $\pm$ (16.94\%)} & \textbf{0.7487 $\pm$ (0.1256)} \\ \hline
\end{tabular}%
}
\end{table*}

\subsubsection{Bayesian analysis}
\label{subsec:results:bayesiananalysis}
Bayesian correlated t-tests \citep{benavoliTimeChangeTutorial2017} were performed to test the statistical validity of the difference for evaluation metrics used in this study. It evaluates the mean difference of evaluation metrics produced by cross-validation on a single dataset between two models. We consider two models to be practically equivalent when the mean difference of AUC and H measure is less than $0.5\%$, and the mean difference of KS is less than $0.005$ for a dataset. Since larger and more challenging datasets are considered in this study, a threshold of $0.5\%$ for the AUC and H measure and a threshold of $0.005$ for KS is more appropriate as the difficulty of improving predictive performance increases. Furthermore, due to the recent large scale of the loan portfolio in financial institutions \citep{cornelliFintechBigTech2023}, even a minor improvement can substantially reduce loan losses, leading to an optimised loan portfolio. Thus, the region of practical equivalence (ROPE) is defined as $0.005$. Bayesian correlated t-tests are then used to compare 2D CNN with \textit{Tabular Image} and each benchmark model. Each test produces three posterior probabilities: the posterior probability $P(\textit{2D\; CNN})$ that 2D CNN with \textit{Tabular Image} performs practically better than a benchmark classifier; the posterior probability $P(\textit{ROPE})$ that the two classifiers being practically equivalent; the posterior probability $P(\textit{Benchmark})$ that a benchmark classifier performs practically better than the 2D CNN with \textit{Tabular Image}. We consider the result as significant if one of the three probabilities exceeds $95\%$, along with introducing posterior odds by computing $o(\textit{2D\; CNN},\;\textit{Benchmark}) = P(\textit{2D\; CNN})/P(\textit{Benchmark})$, $o(ROPE, \textit{2D\; CNN}) = P(ROPE)/P(\textit{2D\; CNN})$, and $o(ROPE, \textit{Benchmark}) = P(ROPE)/P(\textit{Benchmark})$ to avoid limited dichotomous thinking when evaluating the results \citep{gunnarsson_deep_2021}. 
 
\begin{figure}[hbt!]
\centering
\includegraphics[width=\textwidth]{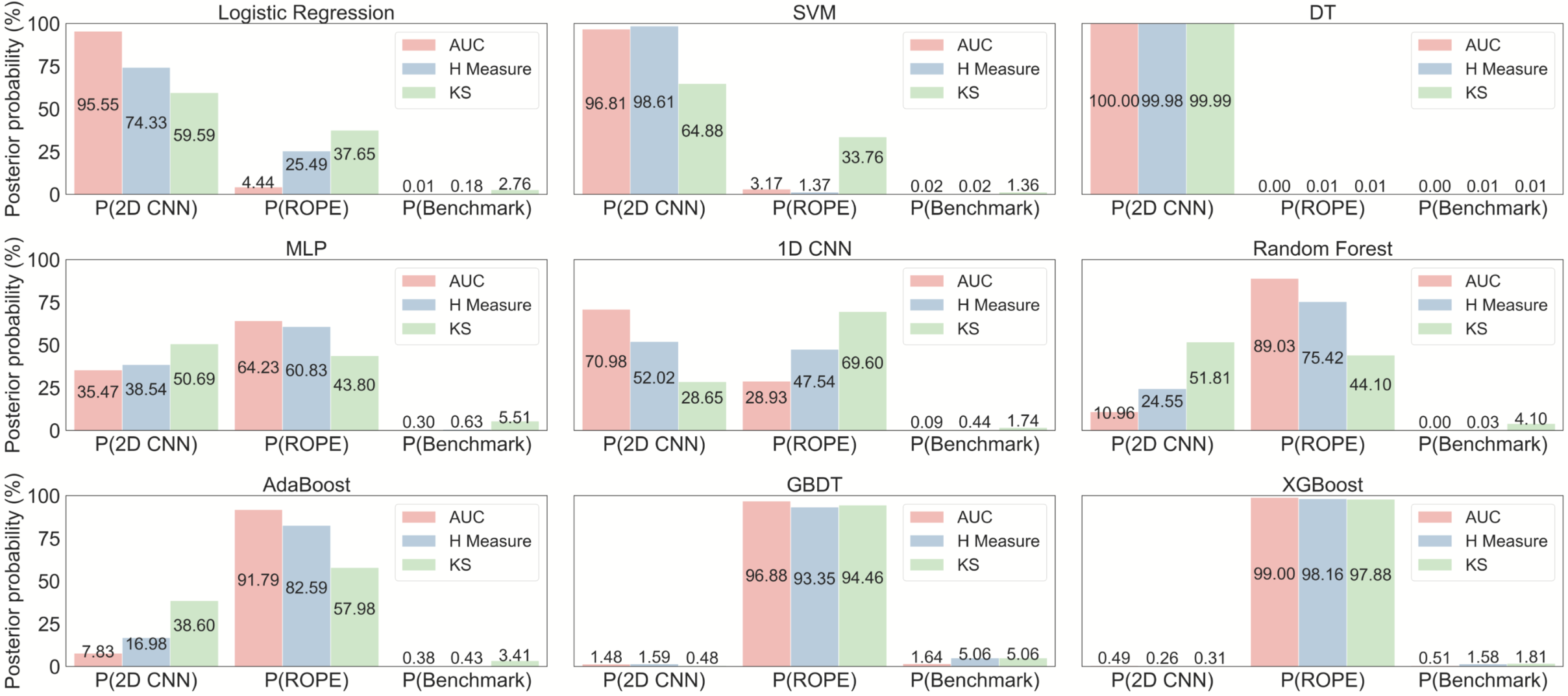}
\caption{Bayesian correlated t-tests for the difference between the 2D CNN with \textit{Tabular Image} and each classifier considered in this study for each performance metric for the TC dataset. It shows a $3\times3$ matrix of bar plots. Each bar plot has three types of bars:$P(\textit{2D\; CNN})$ is the posterior probability that 2D CNN with \textit{Tabular Image} performs practically better than the classifier mentioned in the title of each bar plot; $P(\textit{ROPE})$ is the posterior probability of the two classifiers being practically equivalent; and $P(\textit{Benchmark})$ is the posterior probability that the classifier mentioned in the title of each bar plot performs practically better than the 2D CNN with \textit{Tabular Image}. The number on each bar represents the posterior probability of the Bayesian correlated t-test for each metric.}
\label{fig:bayesian_taiwan}
\end{figure}

Figure \ref{fig:bayesian_taiwan} shows the results for 2D CNN with \textit{Tabular Image} compared with each of the benchmark models for the TC dataset. For individual classifiers, as shown in Figure \ref{fig:bayesian_taiwan}, 2D CNN with \textit{Tabular Image} significantly outperformed Logistic regression, SVM, DT, and 1D CNN with a probability between $95.82\%$ and $100\%$. There is also strong evidence to suggest that 2D CNN with \textit{Tabular Image} performed practically better than MLP based on all three metrics considered with a posterior odd of $o(\textit{2D\; CNN},\;\textit{Benchmark})$ between $6.9$ and $94.3$. Regarding ensemble models, 2D CNN with \textit{Tabular Image} performed significantly better than the random forest in terms of AUC with a probability of $98.19\%$. In addition, strong evidence suggests that 2D CNN with \textit{Tabular Image} performed practically better than AdaBoost based on the AUC ($o(\textit{2D\; CNN},\;\textit{Benchmark})=32.1$) and H-measure ($o(\textit{2D\; CNN},\;\textit{Benchmark})=44.6$). Furthermore, 2D CNN with \textit{Tabular Image} performed equally to the state-of-the-art GBDT and XGBoost based on all three metrics considered in support of positive $o(ROPE, \textit{2D\; CNN})$ ranging from $9.8$ to $56.2$ and positive $o(ROPE, \textit{Benchmark})$ ranging from $1.83$ to $21.5$. 
 
\begin{figure}[hbt!]
\centering
\includegraphics[width=\textwidth]{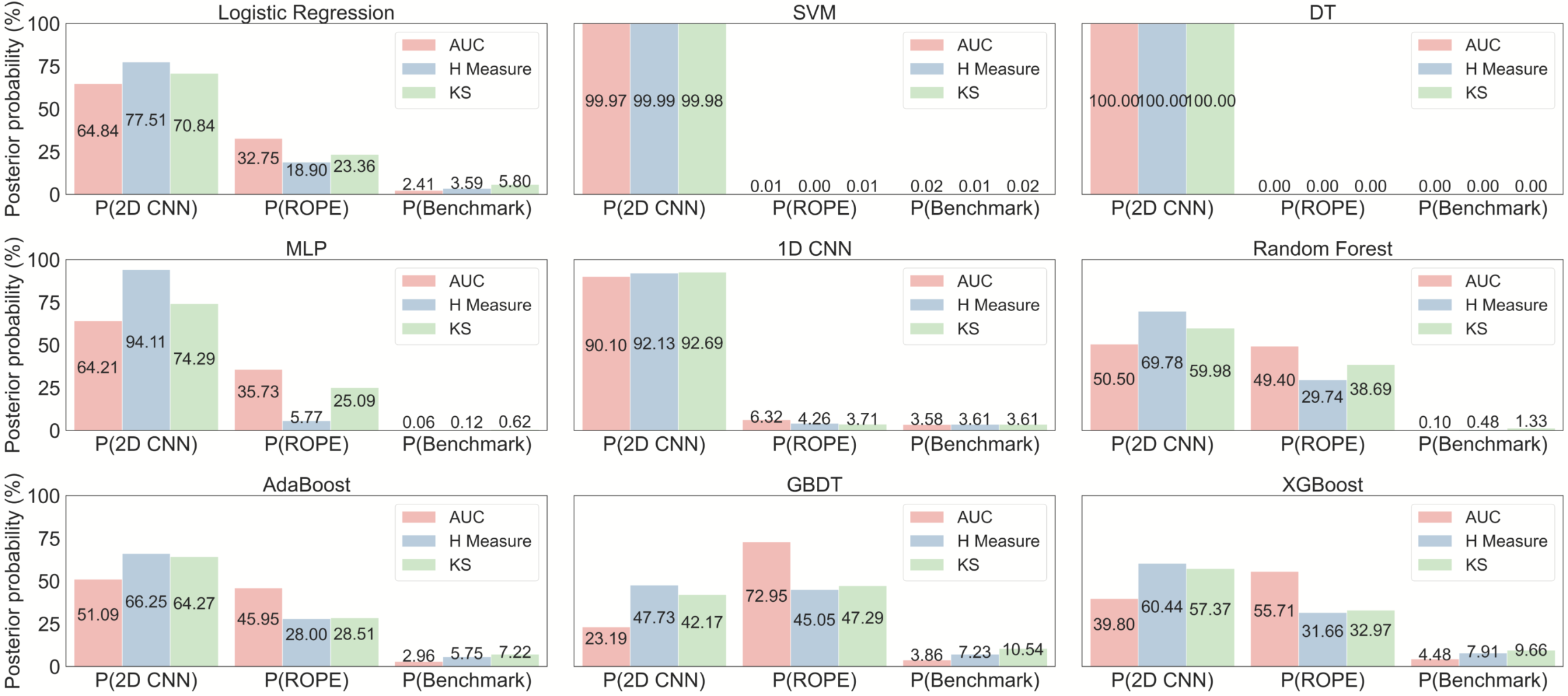}
\caption{Bayesian correlated t-tests for the difference between the 2D CNN with \textit{Tabular Image} and each classifier considered in this study for each performance metric for the HC dataset. It shows a $3\times3$ matrix of bar plots. Each bar plot has three types of bars:$P(\textit{2D\; CNN})$ is the posterior probability that 2D CNN with \textit{Tabular Image} performs practically better than the classifier mentioned in the title of each bar plot; $P(\textit{ROPE})$ is the posterior probability of the two classifiers being practically equivalent; and $P(\textit{Benchmark})$ is the posterior probability that the classifier mentioned in the title of each bar plot performs practically better than the 2D CNN with \textit{Tabular Image}. The number on each bar represents the posterior probability of the Bayesian correlated t-test for each metric.}
\label{fig:bayesian_home}
\end{figure}

Figure \ref{fig:bayesian_home} shows the results for 2D CNN with \textit{Tabular Image} compared with each of the benchmark models for the HC dataset. From the graph, we can see that 2D CNN with \textit{Tabular Image} significantly performed better than SVM, DT, and MLP in terms of AUC and H-measure in support of $\textit{2D\; CNN}$ ranging from $95.99\%$ to $100\%$ and from $98.33\%$ to $100\%$, respectively. In addition, posterior probabilities also support that 2D CNN with \textit{Tabular Image} performed better than Logistic regression, 1D CNN and an ensemble model, random forest, in terms of AUC ($81.53\%-92.77\%$), H-measure ($85.79\%-93.56\%$), and KS ($80.14\%-93.9\%$). It is apparent from this graph that our method positively outperformed all the ensemble boosting models considered, including the state-of-the-art XGBoost, with posterior odds between $3.3$ and $12.1$ for all three metrics. 
 
\begin{figure}[hbt!]
\centering
\includegraphics[width=\textwidth]{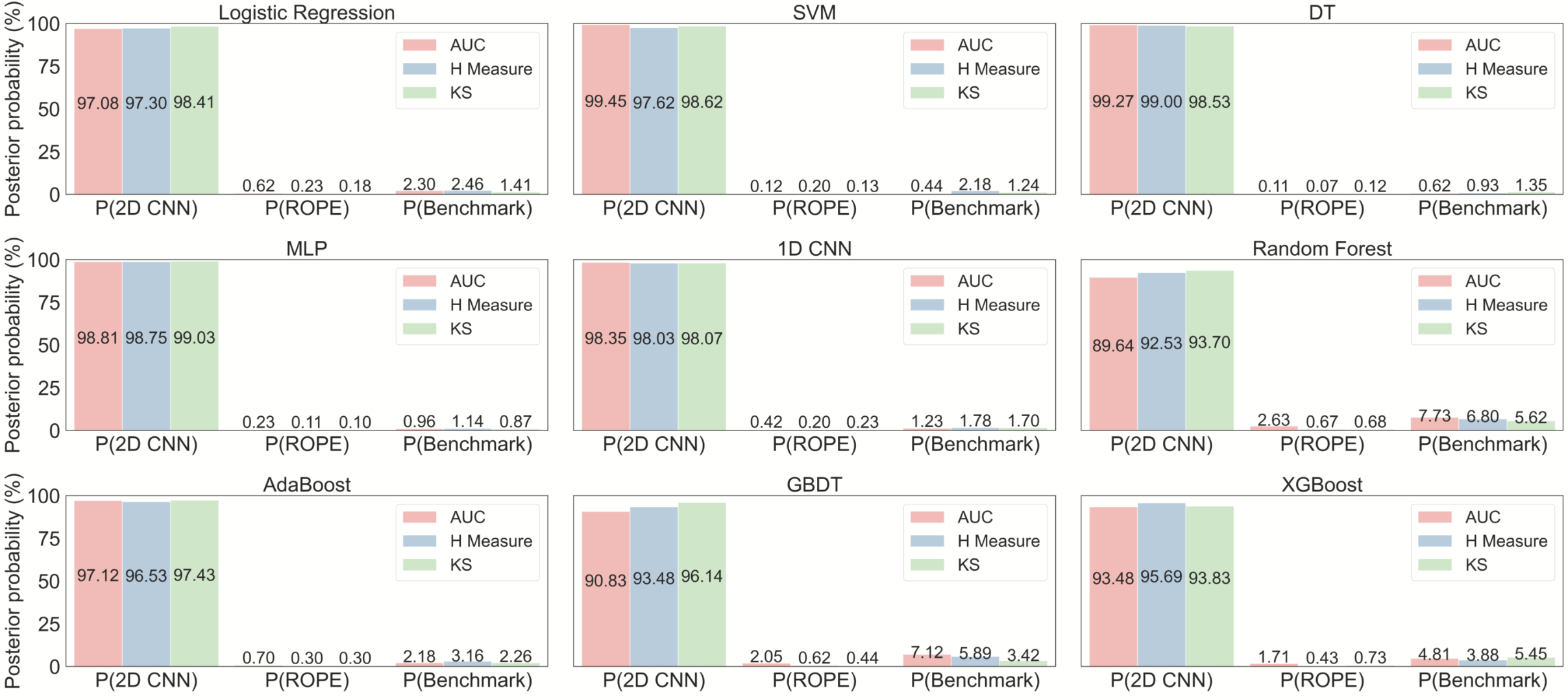}
\caption{Bayesian correlated t-tests for the difference between the 2D CNN with \textit{Tabular Image} and each classifier considered in this study for each performance metric for the FM dataset. It shows a $3\times3$ matrix of bar plots. Each bar plot has three types of bars:$P(\textit{2D\; CNN})$ is the posterior probability that 2D CNN with \textit{Tabular Image} performs practically better than the classifier mentioned in the title of each bar plot; $P(\textit{ROPE})$ is the posterior probability of the two classifiers being practically equivalent; and $P(\textit{Benchmark})$ is the posterior probability that the classifier mentioned in the title of each bar plot performs practically better than the 2D CNN with \textit{Tabular Image}. The number on each bar represents the posterior probability of the Bayesian correlated t-test for each metric.}
\label{fig:bayesian_fannie}
\end{figure}

Figure \ref{fig:bayesian_fannie} shows the results for 2D CNN with \textit{Tabular Image} compared with each of the benchmark models for the FM dataset. From the graph, we can see that 2D CNN with \textit{Tabular Image} significantly performs better than all individual models with a probability between $99.88\%$ and $100\%$. Regarding the ensemble models, random forest and AdaBoost performed significantly worse than 2D CNN with \textit{Tabular Image} in support of a probability between $95.37\%$ and $97.21\%$. In addition, 2D CNN with \textit{Tabular Image}, again, positively outperformed the state-of-the-art GBDT and XGBoost, with posterior odds between $2.3$ and $4.3$ for all three metrics.

\subsection{Comparison of different tabular data-image transformation methods}
\label{subsec:results:comparewithdiffimagetransmethod}
We compared our \textit{Tabular Image} with other popular tabular data-image transformation methods: One-hot \citep{dastile_making_2021}, and DeepInsight \citep{sharma_deepinsight_2019}. Table \ref{table:imagetransresults} shows the average test set AUC, H-measure, KS, and their corresponding standard deviation of five-fold cross-validation for each tabular data-image transformation method with the ConvNeXt on each dataset. 

Overall, our Tabular Images tend to perform better than DeepInsight and One-hot transformation for all evaluation metrics considered in both datasets. Our method outperformed DeepInsight with an improvement in AUC ranging from $11.33\%$ to $25.24\%$ while identifying an increment ranging from $0.89\%$ to $41.75\%$ in AUC compared to the One-hot transformation. Compared to One-hot and DeepInsight, our proposed \textit{Tabular Image} can better preserve information than data transformation methods like DeepInsight and One-hot, which apply dimension reduction techniques or use a sparse matrix with value $0$ and $1$ to obtain a 2D image. The use of dimension reduction or a sparse matrix may result in information loss and thus hinder performance. Interestingly, we can see that the AUC of DeepInsight in TC and HC and the AUC of One-hot in the FM dataset are only slightly above $50\%$, showing limited improvement compared to random guesses. These results show that the performance of One-hot and DeepInsight is unstable across three datasets that have varying scales and complexity. In contrast, \textit{Tabular Image} offers a more informative and stable approach that yields state-of-the-art performances across datasets of varying scales and complexity.

\begin{table*}[hbt!]
\centering
\caption{The average (standard deviation) test set AUC and KS of five-fold cross-validation for different image transformation methods on each dataset.}
\label{table:imagetransresults}
\resizebox{\textwidth}{!}{%
\begin{tabular}{lllll}
\hline
Dataset & Data transformation method & AUC & H measure & KS \\ \hline
TC & DeepInsight & 52.74\% $\pm$ (12.32\%) & 5.61\% $\pm$ (4.61\%) & 0.2225 $\pm$ (0.0869) \\
 & OneHot & 77.09\% $\pm$ (0.90\%) & 28.32\% $\pm$ (2.09\%) & 0.4236 $\pm$ (0.0238) \\
 & Tabular Image & \textbf{77.98\% $\pm$ (1.03\%)} & \textbf{28.88\% $\pm$ (2.08\%)} & \textbf{0.4281 $\pm$ (0.0236)} \\
 &  &  &  &  \\
HC & DeepInsight & 53.84\% $\pm$ (6.31\%) & 1.41\% $\pm$ (1.53\%) & 0.0951 $\pm$ (0.0346) \\
 & OneHot & 72.43\% $\pm$ (0.41\%) & 15.72\% $\pm$ (0.70\%) & 0.3348 $\pm$ (0.0073) \\
 & Tabular Image & \textbf{75.06\% $\pm$ (0.36\%)} & \textbf{19.70\% $\pm$ (0.72\%)} & \textbf{0.3729 $\pm$ (0.0093)} \\
 &  &  &  &  \\
FM & DeepInsight & 80.42\% $\pm$ (6.35\%) & 29.72\% $\pm$ (10.07\%) & 0.4941 $\pm$ (0.1031) \\
 & OneHot & 50.00\% $\pm$ (0.00\%) & 0.00\% $\pm$ (0.00\%) & 0.0000 $\pm$ (0.0000) \\
 & Tabular Image & \textbf{91.75\% $\pm$ (1.15\%)} & \textbf{55.05\% $\pm$ (3.21\%)} & \textbf{0.6827 $\pm$ (0.0355)} \\ \hline
\end{tabular}%
}
\end{table*}

\subsection{Robustness across different image sizes on Tabular Image}
\label{subsec:results:comparewithdiffimagesize}

\begin{figure}[h!]
    \centering
    \includegraphics[width=\textwidth]{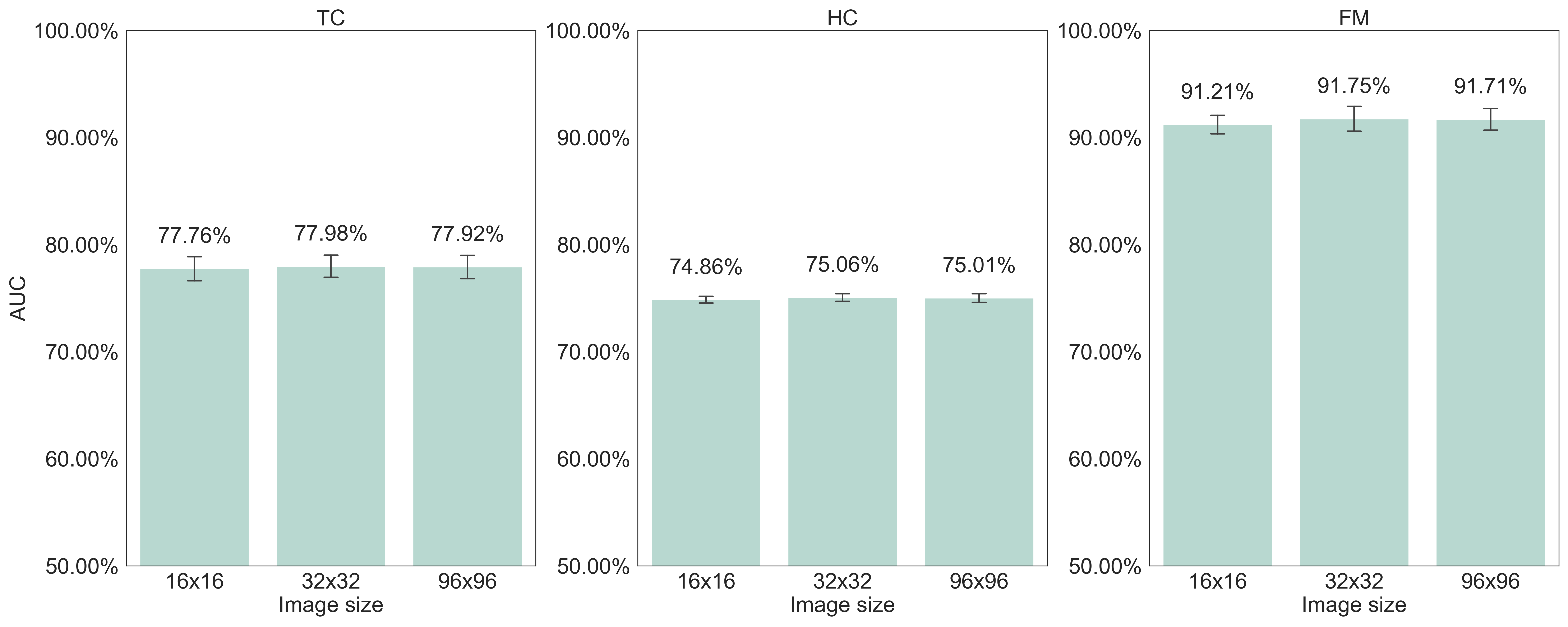}
    \caption{The average AUC across three different image sizes in the five-fold cross-validation.}
    \label{fig:diffImageSizeBarChart}
\end{figure}

To test whether our proposed method is robust to the image size, we ran \textit{Tabular Image} with different image size parameters on the TC, HC, and FM datasets. We ran the \textit{Tabular Image} with three image sizes, small, regular, and large: $16\times16$, $32\times32$, and $96\times96$ pixels, respectively. After data transformation, images of each size were fed into a ConvNeXt model. The optimiser and other hyper-parameters used in the robustness check are consistent with those we used in the section \ref{subsec:experiments:models}. We evaluated the overall discriminative performance by using AUC. Figure \ref{fig:diffImageSizeBarChart} shows the average of AUC in five-fold cross-validation across three image sizes. As can be seen, image sizes of $32\times32$ and $96\times96$ result in stable and similar AUC. However, a slight drop of about $0.2\%$ in AUC can be seen in the TC and HC datasets, and a drop of about $0.5\%$ in AUC can be seen in the FM dataset when the image size shrinks to $16\times16$. A possible explanation for this might be that the model architecture we used in this study is tailored for images greater than or equal to $32\times32$. Therefore, using an image size of $16\times16$ requires resizing the images to at least $32\times32$ using interpolation techniques. The drop in AUC indicates that interpolation techniques may hamper the information in tabular images, making extracting useful information in input images difficult. As a result, we suggest that when scaling tabular images, one should try to adjust the image size to the suggested image size of the selected 2D CNN when using \textit{Tabular Image}. As a resilient method, \textit{Tabular Image} can adjust the size of images without losing any information. This means that as long as the image size is sufficient for the input features one plans to use, it can be shrunk or enlarged to match the 2D CNN requirements, making it easier to use advanced 2D CNNs with different input size requirements with minimal variance of performance.

\subsection{Robustness across different feature arrangement methods on Tabular Image}
\label{subsec:results:comparewithdiffimageallocation}
\begin{figure}[h!]
    \centering
    \includegraphics[width=\textwidth]{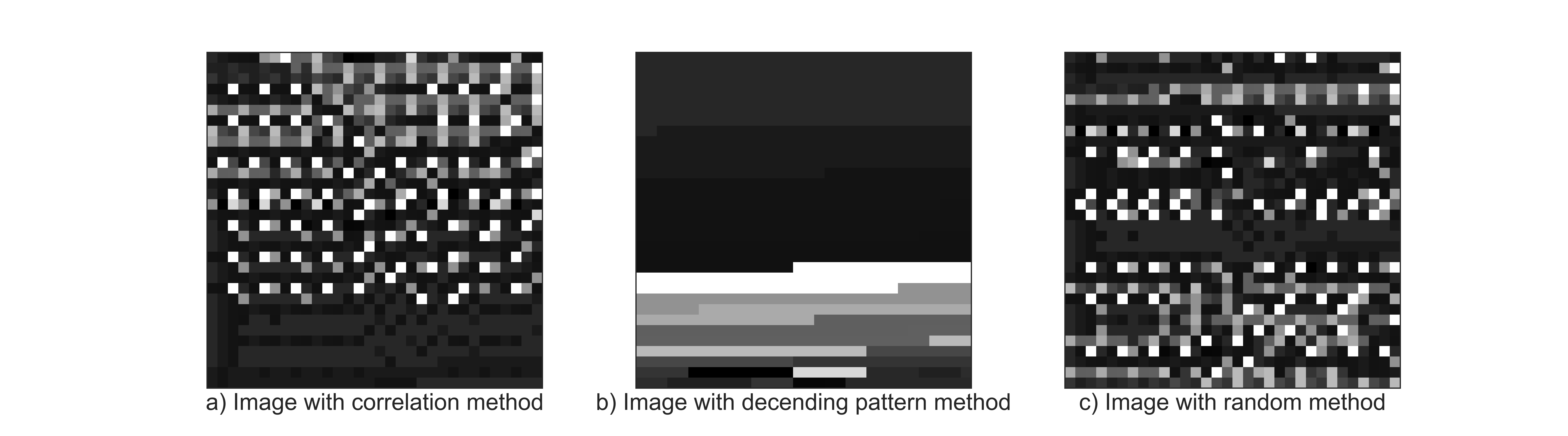}
    \caption{Example images of three types of feature arrangement methods of a sample in the TC dataset. These three images were generated following the correlation method, descending pattern method, and random method, respectively.}
    \label{fig:diffPixelArrExamples}
\end{figure}

\begin{figure}[h!]
    \centering
    \includegraphics[width=\textwidth]{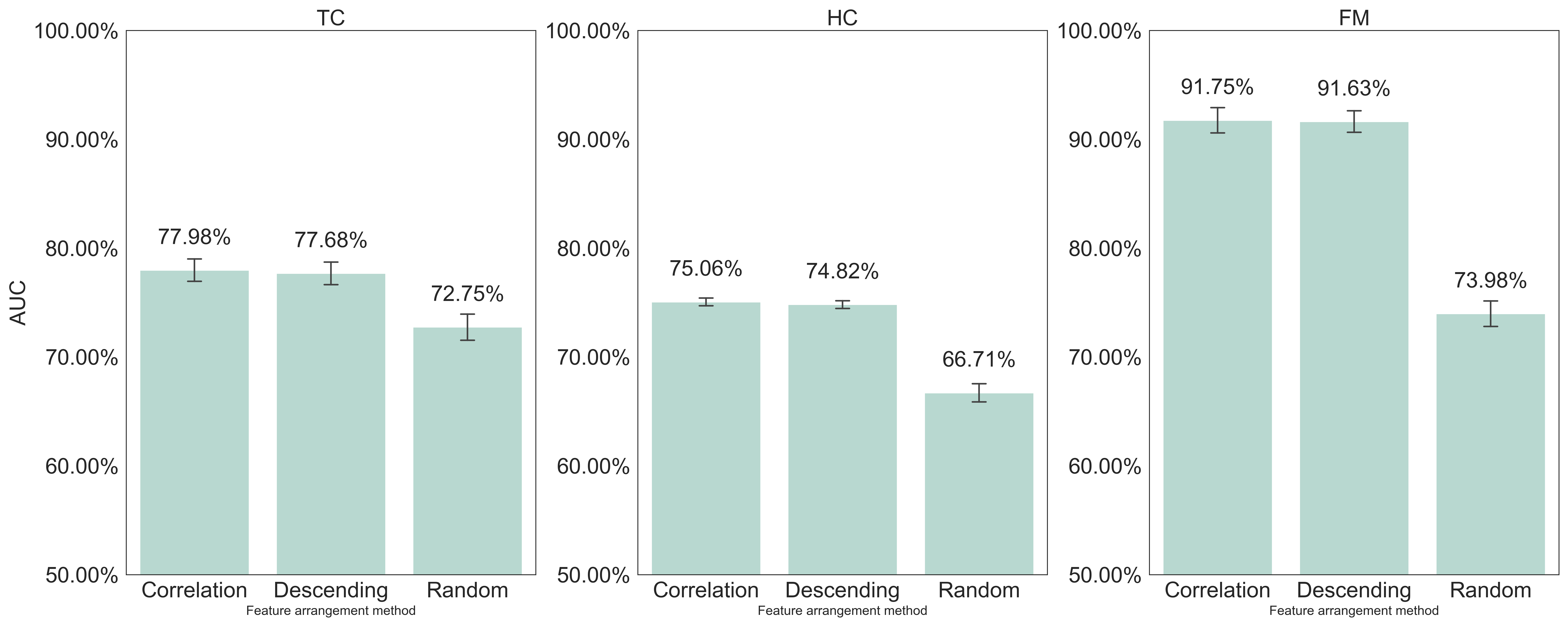}
    \caption{The average AUC of feature arrangement methods in the five-fold cross-validation in the TC dataset.}
    \label{fig:diff_feature_arrangement}
\end{figure}

\begin{figure}[hbt!]
\centering
\includegraphics[width=\textwidth]{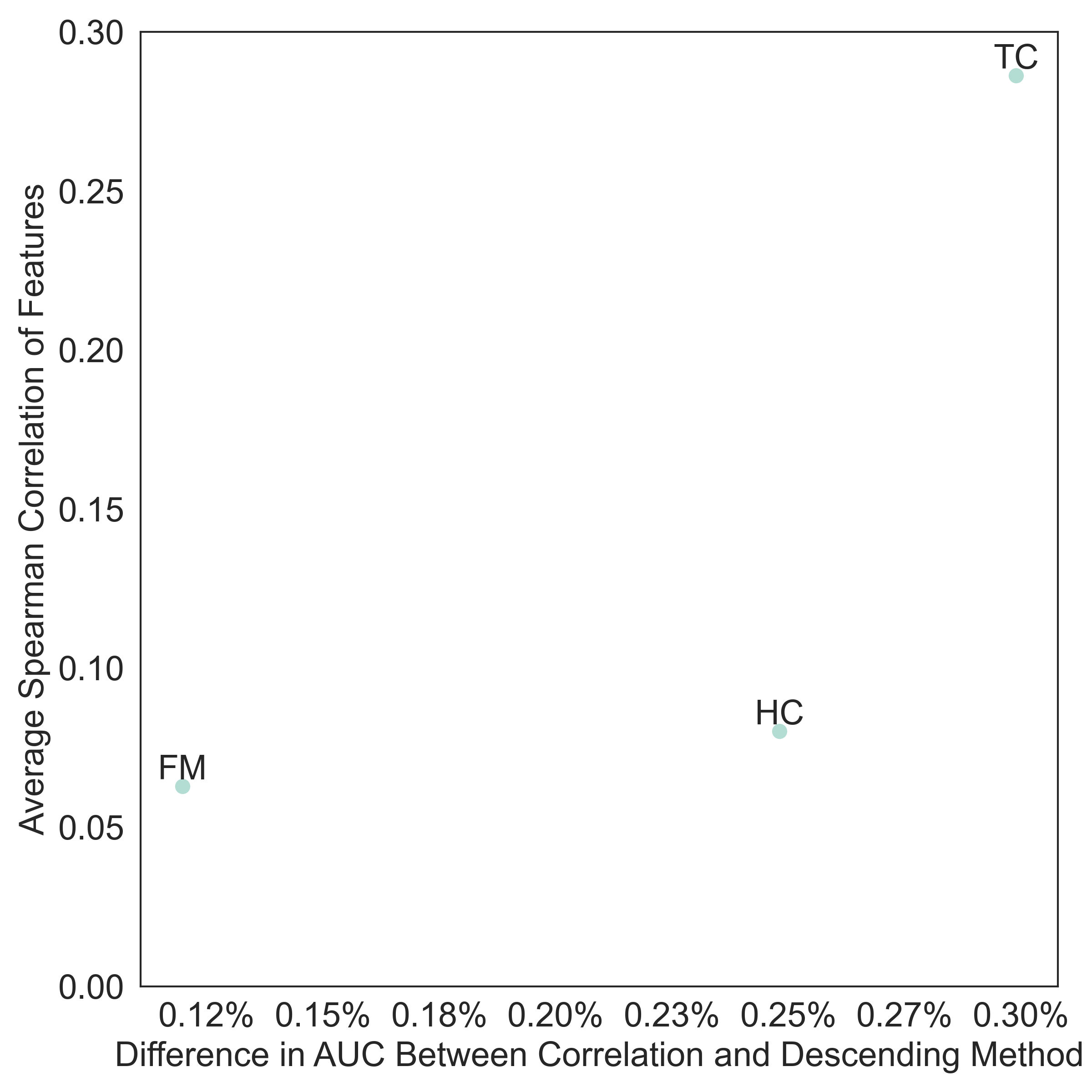}
\caption{Relationship between Average Spearman Correlation of Features and the Difference in AUC Between Correlation and Descending Methods for TC, HC, and FM}
\label{fig:diff_in_auc_corr_desc}
\end{figure}

We also ran the \textit{Tabular Image} with three different feature arrangement methods, namely correlation method, descending pattern method, and random ordering, on the TC, HC and FM datasets to test the robustness of our proposed method to feature arrangement methods. Figure \ref{fig:diffPixelArrExamples} shows the visualisation of a sample's three different feature arrangement methods in the TC dataset. We first generated tabular images using the correlation method, as described in section \ref{subsec:featureArr}. The descending pattern method is similar to the correlation method, except that the order of the feature arrangement matrix is decided by the descending order of the IV of the corresponding feature rather than a correlation coefficient matrix. In other words, values of the same feature are adjacent to each other, which minimises the interaction among features. Finally, we also generated tabular images using a random method by randomly shuffling the pixel locations for each image. In contrast to the aforementioned two methods that follow specific patterns, the pixel locations of every image in the random method do not follow the same pattern. In other words, each image follows a different feature arrangement pattern in the random method. The image size of three different feature arrangement methods was fixed to $32\times32$ pixels for a fair comparison. Next, each type of image was used as input in a ConvNeXt model, and the prediction performance was evaluated using AUC in five-fold cross-validation. Figure \ref{fig:diff_feature_arrangement} shows the average AUC in five-fold cross-validation across different feature arrangement methods. The AUC of the correlation method outperformed the descending pattern method, highlighting that correlation among features should be considered and can improve model performance when applying 2D CNN. In addition, we explored the relationship between the average Spearman correlation of features and the difference in AUC between correlation and descending Methods for TC, HC, and FM. From Figure \ref{fig:diff_in_auc_corr_desc}, it can be seen that as the average Spearman correlation of features becomes stronger, the AUC between correlation and descending Methods increases, indicating \textit{Tabular Image} can effectively extract information from Spearman correlation and improve the predictive performance. Furthermore, it is worth noting that the AUC of the random method significantly dropped from $5.23\%$ to $17.79\%$ compared to the correlation method, demonstrating that all images should follow the same pattern other than randomly arranging each image; otherwise, it is difficult for 2D CNNs to extract relationships between input features and default risk.

\subsection{Robustness of random oversampling on Tabular Image}
\label{subsec:results:comparewithros}
\begin{figure}[hbt!]
\centering
\includegraphics[width=\textwidth]{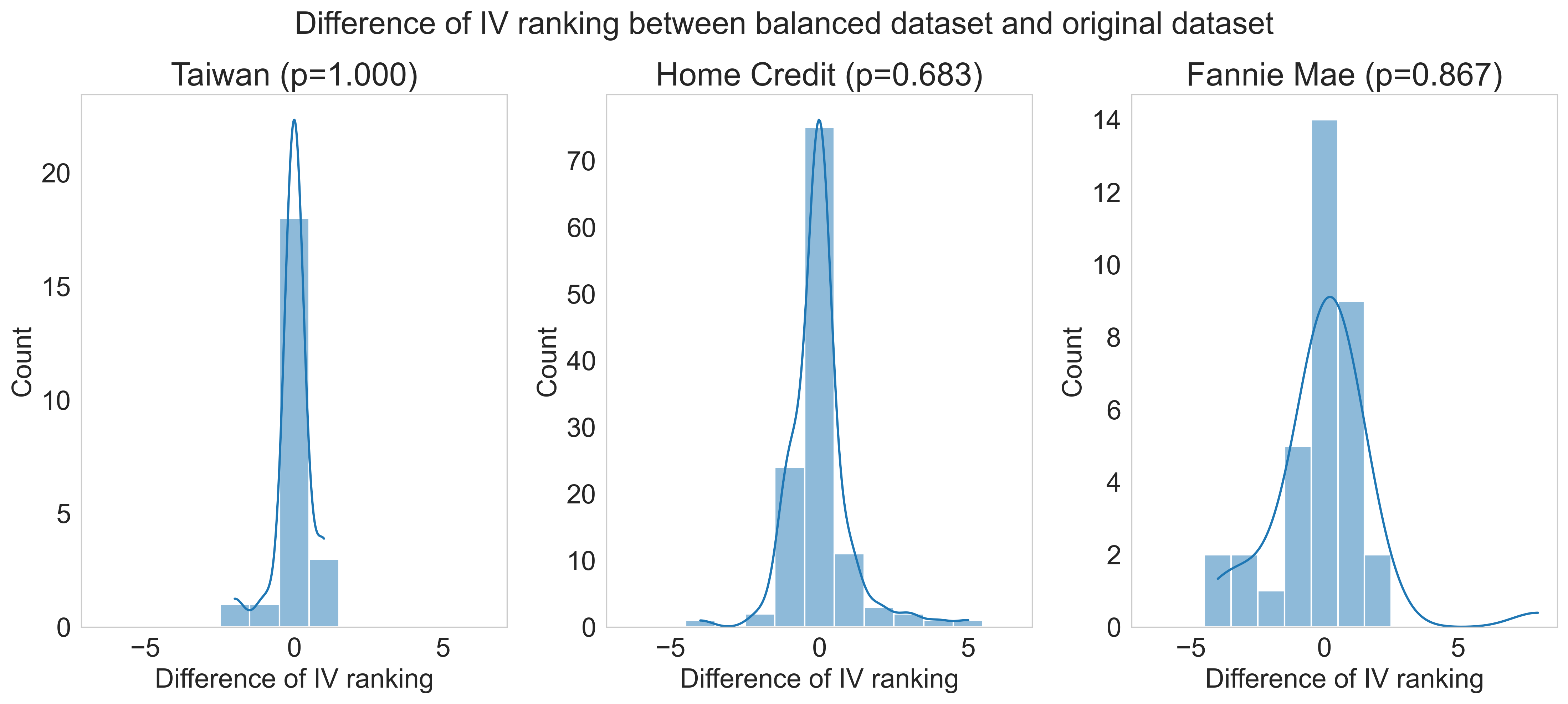}
\caption{Distribution of the difference in Information Value (IV) ranking between balanced datasets and original datasets across three different datasets: Taiwan, Home, and Fannie Mae. The p-values, derived from the Wilcoxon signed-rank test, indicate the significance level of the differences observed.}
\label{fig:iv_ranking_diff}
\end{figure}

Class imbalance is a significant challenge in credit scoring applications, impacting the IV and its ability to evaluate the importance of features. As we use random oversampling as a rebalancing technique to mitigate the class imbalance problem, experiments are conducted to assess whether the ranks of IVs of features changed before and after applying random oversampling across the three datasets used in this study. The changes are calculated by subtracting the rank of IV before applying random oversampling from the rank of IV after applying random oversampling. Figure \ref{fig:iv_ranking_diff} shows the histograms to illustrate the distribution of IV ranking differences, where most differences cluster around zero, indicating the minimal impact of class balancing on the ranks of IVs. The p-values, derived from the Wilcoxon signed-rank test, with the Taiwan Credit ($p=1.000$), Home Credit ($p=0.683$), and Fannie Mae ($p=0.867$) showing no significant difference between the ranks of IVs of features before and after applying random oversampling. This suggests that our method is stable when using random oversampling, as random oversampling does not significantly alter the ranks of features as measured by IV.

\section{Summary and Discussion}
\label{sec:dis_concl}


We set out to take advantage of 2D CNNs in credit scoring tasks and mitigate the challenges neural networks encountered when applied to tabular data. The present research developed the \textit{Tabular Image}, a novel data transformation method to convert tabular data into images for 2D CNNs and mitigates the mixed feature type problem and the data sparsity problem for credit scoring tasks. To investigate its effectiveness, we applied the \textit{Tabular Image} to three credit datasets: TC, HC and FM. We trained a 2D CNN with tabular images to predict the probability of default on credit applicants. To better evaluate the proposed \textit{Tabular Image}, the prediction performances were compared with nine benchmark machine learning methods and two tabular data-image transformation methods. We also evaluated the robustness of the proposed \textit{Tabular Image}. 

The comparison with machine learning methods showed that the performance of 2D CNN trained on tabular images is consistently good, and its prediction advantage is more evident in the more complicated dataset, confirmed by Bayesian correlated t-tests. Particularly, our proposed method with a deep 2D CNN constantly outperformed its shallower counterpart, 1D CNN. These results indicate that it is possible to dig more useful signals for default prediction utilising 2D CNNs with a proper data representation method. This finding is especially valuable for companies which suffer from severe information asymmetry. For instance, companies serving subprime borrowers with little or no credit history usually face a high delinquency rate as borrowers tend to have higher credit risk. Meanwhile, the lack of credit data prevents these financial institutions from building better credit scoring models from existing data. Therefore, even a small fraction of improvement using existing data can be considered significant, saving a huge amount of loan loss. The results also show that the performance of 2D CNN with tabular images continues to increase as the sample size increases, which is consistent with the results of \citet{doumpos_operational_2023, grinsztajnWhyTreebasedModels2022a}, indicating a promising future of applying \textit{Tabular Image} to big data in credit scoring tasks.

The comparison with two tabular data-image transformation methods showed that \textit{Tabular Image} provides the best prediction performance among the data transformations used in this study. Besides prediction performance, \textit{Tabular Image} shows advantages compared to existing tabular data-image transformation methods. First, \textit{Tabular Image} creates compact images while the DeepInsight and One-hot transformation method creates images where a large proportion of the matrices are blank (i.e. $0$). Moreover, for the One-hot transformation method, the number of blank parts will increase if there are more bins in a discretised feature or if a categorical feature has many unique values. Second, \textit{Tabular Image} can mitigate information loss by directly using feature values as pixels compared to DeepInsight, which needs to perform a dimension reduction technique before transforming tabular data to images. Third, our proposed method can create adaptive image width and height to suit the different input requirements of 2D CNNs. In contrast, the image size of the One-hot transformation method is highly related to the number of features and WOE bins. For instance, if one needs to apply deep 2D CNNs that require $224\times224$ image, the dataset should have 224 features with a maximum bin of 224, which is usually difficult to achieve.

The robustness check on different image sizes for \textit{Tabular Image} suggests that the input image size should match the requirement of the 2D CNN planned to use to gain the maximum performance. When using the proposed \textit{Tabular Image} method, changing the image size will not affect the pixel values or the proportion of features in the image. Therefore, as long as the pixel values and number of features remain consistent and the image size is large enough to include all planned features, it should be possible to adjust the image size to meet the input requirements of different 2D CNN architectures without losing any information. This observation may be crucial to tabular data-image transformation methods applied to credit data because the traditional interpolation techniques for images may introduce unexpected noise when scaling an image, turning minority samples into majority samples and vice versa, damaging the dataset and misleading the model training process. In addition, this study investigated the performance of different feature arrangement methods applied to \textit{Tabular Image}. We demonstrated that spatial relationships should be considered while forming images. This study used the Spearman correlation coefficient matrix to evaluate spatial relationships between features and yield better results than DeepInsight and the One-hot transformation method. We also showed that during transformation, all images need to follow the same feature arrangement method; otherwise, it is difficult for a 2D CNN to extract relationships within a dataset.

The effectiveness of deep learning methods in credit scoring remains a topic of active debate, particularly when compared to the state-of-the-art XGBoost. Our latest model outperformed the XGBoost and other state-of-the-art models in large and challenging datasets. Moreover, we see it as a complementary approach that can enhance predictive performance on large datasets and offer additional benefits. XGBoost performs well on small datasets, but its performance on large datasets is less explored \citep{gunnarsson_deep_2021, lessmann_benchmarking_2015}. On the two larger datasets in our experiments, our method outperformed XGBoost in terms of AUC, H-measure, and KS, with positive results for all three metrics confirmed by Bayesian correlated t-tests. The consistent improvement across multiple large datasets demonstrates the robustness of our approach, showing the potential of applying 2D CNN on large datasets, which aligns with \citet{borisovDeepNeuralNetworks2022}. We further demonstrated, using the HC and FM datasets, that our method outperformed all benchmark models in predicting unbanked borrowers who have limited credit history and subprime borrowers who have lower credit scores. By applying 2D CNNs with the \textit{Tabular Image}, lenders can more accurately assess the risk of unbanked and subprime borrowers, potentially expanding their customer base and increasing profitability. For unbanked and subprime borrowers who usually suffer from high interest rates due to the increased risk, our method can improve access to loans with fair interest rates by correctly estimating the risk, making loans less expensive, and thereby promoting greater financial inclusion. 
  
Besides findings demonstrated by our experiments, our method introduces several additional unique contributions. 
First, \textit{Tabular Image} introduces a novel way of converting tabular data into images by embedding the WOE and IV. This transformation allows us to leverage two-dimensional convolutional neural networks (2D CNNs), which are traditionally used for image data, thereby opening new avenues for applying powerful 2D CNNs to tabular data. 
Second, while XGBoost captures interactions and non-linear relationships through feature splitting, 2D CNNs are capable of extracting new and high-level features by leveraging spatial patterns and relationships between features within images. With the ability of feature extraction, 2D CNN with \textit{Tabular Image} is able to explore and discover useful new features, thus enhancing feature engineering, one of the most important steps in developing a credit scoring model. Furthermore, new features can be used as input for XGBoost to create ensemble frameworks to further improve the performance \citep{khanCNNXGBoostFusionbasedAffective2022, thongsuwanConvXGBNewDeep2021}. 
Third, 2D CNNs can take advantage of the latest developments in both hardware and software, such as advanced GPUs and their corresponding software that allow for faster training processes and efficient training on extensive data, which is beneficial in the real-world lending business as the amount of loan has increased exponentially in the recent decade\footnote{\url{see: https://data.worldbank.org/indicator/FS.AST.PRVT.GD.ZS}}.

The proposed method also contributes to the improvement of explainability, which is a critical requirement in credit scoring, emphasised by regulators across many countries \citep{buckerTransparencyAuditabilityExplainability2022}. To address this requirement, regulators such as the European Banking Authority and the French Prudential Supervision and Resolution Authority recommend using model-agnostic interpretation approaches to achieve interpretability \citep{chenInterpretableMachineLearning2024}. The proposed \textit{Tabular Image} performs a lossless data transformation, preserving original feature information and ensuring compatibility with model-agnostic explainability methods, such as SHapley Additive exPlanations (SHAP)\citep{lundbergUnifiedApproachInterpreting2017}, which have been frequently used to address the explainability problem in credit scoring \citep{chenInterpretableMachineLearning2024,korangiTransformerbasedModelDefault2023, talaatInterpretableCreditScoring2024, zandiAttentionbasedDynamicMultilayer2024}. Because each pixel in the tabular image directly corresponds to a specific feature in the original tabular data, standard explainability techniques that generate pixel-level importance, including SHAP, can be easily and directly applied to explain the predictions made by the downstream 2D CNN models. Such seamless compatibility with established interpretation methods ensures explainable model predictions, aligning closely with regulatory expectations without introducing additional complexity.

Another significant aspect of 2D CNN with \textit{Tabular Image} is its future potential for practical applications. Our method enables transfer learning \citep{alzubaidiReviewDeepLearning2021a}, which XGBoost does not naturally support. 2D CNN models can be pre-trained on large \textit{Tabular Image} datasets and fine-tuned on a small one. By leveraging existing large datasets, transfer learning provides the potential to mitigate the cold-start problem, especially in scenarios when labelled data in the target domain is limited, such as launching a new business or expanding into a new market. In addition, our \textit{Tabular Image} method could potentially enhance traditional visualisation techniques, facilitating a clearer understanding and the identification of unusual patterns within the data. For example, \textit{Tabular Image} can be embedded into the loan approval process to provide intuitive images for loan officers. This makes it more efficient to initially identify the risk of a potential borrower based on the brightness and pattern of the tabular image before checking various criteria in a tabular format that is not user-friendly. Besides optimising the decision process of loan officers, converting credit data into image format allows for the application of advanced image processing tools and techniques (e.g. pattern recognition algorithms) \citep{paolantiMultidisciplinaryPatternRecognition2020, schmidhuber_deep_2015}, further aiding in the detection of subtle patterns and correlations that might be missed in tabular formats. Additionally, this image-based approach aligns with the trend towards more interactive and user-friendly data analytics tools \citep{keimVisualizationTechniquesMining1996,leiteEVAVisualAnalytics2018,perrotLargeInteractiveVisualization2015}, which are generally more engaging and easier to interpret.

From a managerial perspective, the findings of this research provide insights into the potential of applying 2D CNNs to large credit datasets in real-world applications. On the one hand, because of the large number of total loans today, even a small fragment of performance improvement by applying 2D CNNs with our proposed method can translate into a huge amount of loan loss savings. For instance, according to the 2019 financial report\footnote{\url{https://www.homecredit.net/financial-disclosures/}} of Home Credit Group B.V., the provision for Expected Credit Losses on loans to customers was EUR 1.6 billion as of 31 December 2019, which indicate that even a $0.1\%$ improvement may result in a loan saving of EUR 1.6 million. On the other hand, with more accurate credit scoring models, the possibility of the subprime and unbanked population accessing loans with fair interest rates may increase, thus increasing financial inclusion.
  
Besides promising results in credit scoring tasks, \textit{Tabular Image} offers an adaptable framework that can be expanded to accommodate various datasets in other domains quickly. First, a binning algorithm can be designed and used to select discretisation cut points for numerical features to accommodate the aim of the research. For instance, Chi-squared, tree-based, or entropy-based binning can also be applied to find appropriate cut points. Second, different measures can be implemented to numerically represent the strength of a bin, as long as the measure represents the importance of a feature in the research domain. For instance, entropy can be used to measure the impurity of a bin and information gain can be used to evaluate the importance of a feature. Lastly, the technique to evaluate correlation or distance between features, such as Kendall rank or Euclidean distance, can be changed to the one that better represents the relationship between two features in a specific field. This flexible framework leads to the potential of adjusting \textit{Tabular Image} to suit various tabular datasets and requirements in other domains. For example, the field of credit card fraud detection usually contains features mixed with numerical and categorical features such as transaction amounts, types of cards, merchant information, and digital footprints generated by users, all in a tabular format. Transforming this data into images allows CNNs to capture correlations and subtle anomalies that might be overlooked by traditional methods, enhancing the detection of fraudulent activities and helping both normal users and companies to reduce potential loss. Another example of a potential field of application may be energy consumption forecasting. Energy providers usually collect features such as consumption patterns, weather conditions, and user demographics. These features are usually present in a tabular format and might exhibit high correlations. By classifying energy consumption patterns (e.g., high energy consumption, moderate, or low), energy providers can manage energy more efficiently, increasing environmental sustainability. Tabular data is one of the most common data types in real-world applications and is widely used in applications that are based on relational databases. Our proposed method, therefore, provides the potential to help other domains take advantage of advanced 2D CNNs to improve the model performance in these domains further.

\section{Conclusion and future work}
\label{subsec:contri}

This study has shown that with \textit{Tabular Image}, 2D CNNs can yield good predictive performance in credit scoring. We further extend the work of \citet{gunnarsson_deep_2021} by exploring the possibility of 2D CNNs and testing models on large datasets in the credit scoring field. By proposing \textit{Tabular Image}, this study provides a novel way to convert tabular data into compact images to take advantage of a 2D CNN by embedding two classical pieces of information used in credit scoring tasks, namely WOE and IV, in the image. Through rigorous testing of various models, we demonstrate that the proposed method with a deep 2D CNN exhibits state-of-the-art predictive performance, especially in handling large and complicated datasets while better preserving information in tabular data compared to other tabular data-image transformation methods. This opens a gateway for applying powerful deep 2D CNNs and their corresponding modules, which have already demonstrated impressive performance in other domains and are supported by advanced hardware and software techniques in the credit scoring field. Furthermore, with the flexible framework provided by \textit{Tabular Image}, this advancement can be further extended to various other fields as tabular data is one of the most common data types in real-world applications and is widely used in medicine, finance, manufacturing, fraud detection, and many other applications that are based on relational databases.

Future research could explore more existing modules like advanced optimisers or regulation techniques like label smoothing \citep{mullerWhenDoesLabel2019} to further improve the performance of 2D CNNs. Researchers may also try to explore more advanced setups for 2D CNNs, like masked image modelling \citep{hanRevColV2ExploringDisentangled2023}, or develop specially designed 2D CNNs for tabular image tasks. Future work could also usefully explore how data augmentation techniques \citep{xuComprehensiveSurveyImage2023} in the image recognition domain can mitigate the long-lasting data imbalance problem in credit scoring. In addition, further research should be undertaken to explore how to improve the explainability of images to facilitate the analysis of a single observation to enhance instance-level model explanations. Visual explanation techniques like Grad-CAM \citep{selvarajuGradCAMVisualExplanations2017} and Saliency Maps \citep{simonyanDeepConvolutionalNetworks2014}  could also be investigated to enhance the explainability of 2D CNNs.

\backmatter

\bmhead{Acknowledgements}
Part of this work was undertaken on ARC3 and ARC4, part of the High Performance Computing facilities at the University of Leeds, UK.

\section*{Declarations}

\begin{itemize}
\item Conflict of interest There are no financial or non-financial interests that are directly or indirectly related to the work submitted for publication.

\end{itemize}

\noindent

\bigskip

\bibliography{tabular-image-wo_doi_v2}

\end{document}